\documentclass[fleqn,usenatbib]{mnras}

\usepackage{newtxtext,newtxmath}
\usepackage{soul} 
\usepackage[T1]{fontenc}
\usepackage{ae,aecompl}

\usepackage{graphicx}	
\usepackage{amsmath}	
\usepackage[dvipsnames]{xcolor}

\newcommand{\kms}{\mbox{$\>{\rm km\, s^{-1}}$}}
\newcommand{\pc}{\>{\rm pc}}
\newcommand{\kpc}{\mbox{$\>{\rm kpc}$}} 

\newcommand{\Gyr}{\mbox{$\>{\rm Gyr}$}}
\newcommand{\Myr}{\mbox{$\>{\rm Myr}$}}

\newcommand{\Msun}{\>{\rm M_{\odot}}}

\newcommand\ma{{M1\_c\_b}}
\newcommand\mb{{M2\_c\_nb}}
\newcommand\mc{{M3\_nc\_b}}
\newcommand\md{{M4\_nc\_bd}}

\title[Retrograde stars in disc galaxies]{The relative efficiencies of bars and clumps in driving disc stars to retrograde motion}

\author[Fiteni et al.]{
Karl Fiteni,$^{1}$\thanks{E-mail: karl.fiteni.12@um.edu.mt}
Joseph Caruana,$^{1,2}$
Jo\~ao A. S.  Amarante,$^{3,4}$
Victor P.  Debattista$^{1,5}$ and \newauthor
Leandro {Beraldo e Silva}$^{5}$
\\
\\
$^{1}$Institute of Space Sciences \& Astronomy, University of Malta, Msida MSD 2080, Malta\\
$^{2}$Department of Physics, University of Malta, Msida MSD 2080, Malta\\
$^{3}$Key Laboratory for Research in Galaxies and Cosmology, Shanghai  Astronomical Observatory, \\Chinese Academy of Sciences, 80 Nandan Road, Shanghai 200030, China\\
$^{4}$ University of Chinese Academy of Sciences, No.19A Yuquan Road, Beijing 100049, China \\
$^{5}$Jeremiah Horrocks Institute, University of Central Lancashire, Preston PR1 2HE, UK
}

\date{Accepted XXX. Received YYY; in original form ZZZ}

\pubyear{2021}

\begin{document}
\label{firstpage}
\pagerange{\pageref{firstpage}--\pageref{lastpage}}
\maketitle

\begin{abstract}
The presence of stars on retrograde orbits in disc galaxies is usually attributed to accretion events, both via direct accretion, as well as through the heating of the disc stars. Recent studies have shown that retrograde orbits can also be produced via scattering by dense clumps, which are often present in the early stages of a galaxy's evolution. However, so far it has been unclear whether other internally-driven mechanisms, such as bars, are also capable of driving retrograde motion. Therefore, in this paper, we investigate the efficiencies with which bars and clumps produce retrograde orbits in disc galaxies. We do this by comparing the retrograde fractions and the spatial distributions of the retrograde populations in four $N$-body$+$smooth particle hydrodynamics (SPH) simulations of isolated disc galaxies spanning a range of evolutionary behaviours. We find that both bars and clumps are capable of generating significant retrograde populations of order $\sim 10\%$ of all stars. We also find that while clump-driven retrograde stars may be found at large galactocentric radii, bar-driven retrograde stars remain in the vicinity of the bar, even if the bar dissolves. Consequently, we find that retrograde stars in the Solar Neighbourhood in the clumpy models are exclusively clump-driven, but this is a trace population, constituting $0.01-0.04\%$ of the total stellar population in this region. Finally, we find that neither bars (including dissolving ones) nor clumps in the models are able to produce rotationally supported counter-rotating discs. 
\end{abstract}




\section{Introduction}

According to the prevailing lambda cold dark matter ($\Lambda$CDM) model, the early evolutionary history of disc galaxies was dominated by hierarchical merging and accretion of material \citep{white1991}. However, the subsequent decrease in the frequency of major mergers with time meant that secular, internally driven, processes were able to drive most of the later evolution.

Bars are one of the main drivers of secular evolution in disc galaxies \citep{athanassoula2013}, and previous studies have shown that they can significantly impact their host galaxy through a number of mechanisms (see reviews by \citealt{Kormendy2004, Kormendy2013}). For example, it is well established that bars facilitate the exchange of angular momentum, both throughout the disc, and between different stellar components \citep{hohl1978,sellwood1980}. Angular momentum exchange with the dark matter halo via dynamical friction has also been shown to cause bars to slow down and grow radially \citep{Weinberg1985, Hernquist1992, Debattista2000, Valenzuela2003, valpuesta2006, Chiba2021}. In addition, while bars are generally vertically thin upon forming, they often drive a thickening of the inner disc which leads to the eventual formation of a boxy/peanut bulge \citep{combes1981, combes1990, raha1991,merritt1994, debattista2004, valpuesta2006}. Bars are also found to trigger gas inflows which can lead to the formation of nuclear discs and resonant rings \citep[e.g.][]{ combes1985, buta1996,sakamoto1999,sheth2005,cole+14}. It is also possible for these gas inflows to fuel an active galactic nucleus, although at present there is no consensus on the efficacy of this process (see \citealt{combes2003}, \citealt{ho2008} and references therein). The potential weakening and eventual destruction of bars has also been linked to the build-up of mass in the central regions \citep{hasan1990,pfenniger1990,friedli1993} or to the transfer of angular momentum from the infalling gas to the bar \citep{bournaud2005}, although high resolution simulations have also found the opposite result \citep{shen2004, athanassoula2005, Debattista2006}. In addition, bars are also capable of influencing the overall structure of the disc, for instance through the creation of a break in the radial density profile, which transforms a single component disc into one with a double exponential radial density profile \citep{hohl1971, Debattista2006, Minchev2012, Herpich2017}. There have also been efforts to link the bar with the presence of counter-rotating stellar components. \citet{evans1994} found that stellar counter-rotation may be induced by a dissolving bar, whereas \citet{wozniak1997} interpreted the `wave pattern' found in the velocity curves of some barred S0 galaxies, within the region of the bar \citep{bettoni1989, bettoni1997, zeilinger2001}, and in simulations \citep{pfenniger1984,sparke1987}, to be the result of stars trapped on $x_4$ (quasi-circular retrograde) orbits by a bar.  Given that bars have such a wide range of consequences for disc galaxies, we are motivated to investigate the degree to which they are capable of producing retrograde disc stars, and the implications this might have for disc galaxies such as the Milky Way (MW).

Dense, star-forming clumps in a young disc can also produce retrograde stars \citep[e.g.][]{amarante2020}. High-redshift ($z > 1$) galaxies often display a light distribution dominated by gas-rich, star-forming clumps, giving them an irregular, clumpy appearance. This contrasts with the smoother light distribution usually observed in massive disc galaxies in the local Universe. Clumps were first observed in high-redshift galaxies by \citet{cowie1995} and \citet{van1996}. The clumpy nature of early galaxies was confirmed by more recent observations, owing primarily to instruments capable of deep, high-resolution observations, such as {\it Wide Field Camera 3} (WFC3) and the {\it Near Infrared Camera and Multi-Object Spectrometer} (NICMOS) on board the {\it Hubble Space Telescope}, which probe the rest-frame UV and near infrared regions at z > 1 \citep[e.g.][]{elmegreen2007,elmegreen2008,ravindranath2006,schreiber2011,genzel2011, overzier2010, swinbank2010, guo2012}. Their formation is likely due to gravitational instabilities in the proto-disc, a scenario which is supported by numerical studies \citep{noguchi1999, ceverino2010}. \citet{guo2015} investigated the clump demographics of 3239 galaxies at redshift $0.5 < z < 3.5$ in the CANDELS/GOODS-S and UDS fields and found that  $\approx 55\%$ of intermediate to high-mass galaxies in their sample contain clumps. However, this fraction decreases at lower redshifts to $\approx 15\%$ at $z = 0.5$. On the other hand, $\approx 60\%$ of lower mass galaxies have clumps, with this fraction remaining roughly constant with redshift. The sizes and masses of clumps are still somewhat uncertain, with $\sim 1$ kpc resolution observations yielding clump masses up to $10^{9.5}{\mathrm M_{\odot}}$  \citep{schreiber2011,soto2017}. However, higher resolution studies have demonstrated that the derived masses and linear sizes of clumps are highly dependent on the resolution and sensitivity of observations, and are often systematically overestimated, with masses likely being closer to $\sim 10^{7}-10^{8} \mathrm M_{\odot}$ and linear sizes between $100-500$ pc  \citep{dessauges2017, cava2018}.

Significant work has also gone into understanding the impact that clumps have on the overall evolution of disc galaxies. For example, some studies find that if individual clumps are not disrupted by star formation and live long enough, they sink into the centre of the galaxy, forming a bulge component \citep{elmegreen2007,dekel2009}. More recently, clumps have also been proposed to be the driving mechanism behind the origin of the geometric and chemical thick discs \citep{bournaud2009,Clarke+19, BeraldoeSilva+20}. In addition, using an isolated $N$-body+SPH model similar to the Milky Way, \citet{amarante2020} demonstrated that low angular momentum ($v_{\phi}<100 \kms$) stars, including retrograde ones, are present in the Solar Neighbourhood due to clump scattering in the first Gyrs of evolution, rather than exclusively because of the action of merger events, such as the {\it Gaia}-Enceladus-Sausage \citep{helmi2018, belokurov2018}, as previously suggested. The effect which clumps have on the evolution of galaxies such as the MW is not settled, but it remains possible that they may have played a significant role. Therefore, we also explore the role clumps play in driving retrograde motion in MW-like disc galaxies.

In this paper we make a distinction between retrograde and counter-rotating motion: while both retrograde and counter-rotating stellar populations have the common characteristic of orbiting the host galaxy in an opposite sense to the main stellar disc, we emphasize that the retrograde populations we will explore do not take the form of a rapid counter-rotating disc such as those observed in NGC 7217 \citep{merrifield1994} and NGC 4550 \citep{rubin1992,rix1992}. Counter-rotating discs occur when a substantial population of stars orbit the galaxy with a total angular momentum vector pointed in the opposite direction to that of the main stellar disc, and close to the circular velocity of the system.  Counter-rotation is generally thought to be linked with the accretion of counter-rotating gas which subsequently forms stars \citep[e.g.][]{katkov2013, pizzella2018}. Given the right conditions, a counter-rotating disc may also be produced by mergers \citep[e.g.][]{puerari2001}. On the other hand, the retrograde populations we study in this paper do not rotate as rapidly, and indeed may or may not manifest as a disc, but may be more spheroidally distributed. Merger events have also been shown to heat orbits in a pre-existing disc and produce retrograde populations \citep[e.g.][]{toth1992,velazquez1999,kazandzitis2009,moetezadian2016,Belokurov2019, grand2020}. However, the possibility of internal mechanisms such as the bar or stellar clumps being responsible for driving retrograde motion has not yet been fully explored.

This paper is organized as follows. Section \ref{sims} presents the simulations used in our analysis. In Section \ref{results} we compare the models and show the properties of the retrograde populations in them. In Section \ref{conc}, we discuss the implications of our results for the MW, and for disc galaxies in general. Section \ref{summ} presents a summary of our conclusions.

\section{The Simulations}\label{sims}

We consider four $N$-body+smooth particle hydrodynamics (SPH) simulations (see Table \ref{tab:anysymbols}), spanning a range of evolutionary behaviours which provide us with a mixture of physical mechanisms to study. The initial conditions of all the models embed a gas corona in pressure equilibrium within a co-spatial Navarro-Frenk-White (NFW) \citep{NFW} dark matter halo which constitutes $90\%$ of the mass. All stars form from the gas, with none present in the initial conditions. The first two simulations are {\ma} and {\mb}.  Both of these models formed clumps (hence the "c" in the name), albeit with very different properties.  Model {\ma} formed a large-scale bar (hence the "b" in the name) whereas \mb\ did not ("nb").  These two models have been presented in \citet{ghosh2020} and \citet{BeraldoeSilva+20b}  respectively. Additional properties of these models can be found in those papers. Both models start with identical initial conditions but they are evolved with different sub-grid physics.  The initial conditions are a higher mass resolution version of the model described in \citet{roskar2008}.  

The other two models both form bars; since they start from the same initial conditions, their evolution is very similar at first, until they diverge at later times because they implement different subgrid chemohydrodynamics evolutionary models. In the first one, \mc, the bar is long-lived ("b") while in \md\ the bar that forms subsequently decays ("bd").  Neither of these two models form clumps ("nc").  Model \mc\ has been presented numerous times in previous papers; it is the star-forming model described most extensively in \citet{cole+14} and \citet{Debattista+17}.  Model \md\ was described in \citet{portaluri2017}. The main difference is that in \md, thermal energy and metals can diffuse between gas particles, whereas this is prevented in model \mc. The evolution of the two models is initially quite similar, but their subsequent stochastic evolution sees them diverge when the gas inflow in \md\ overwhelms the bar, leading to its destruction.  By the end of the simulations, a bar of radius $\approx 3 \kpc$ is present in \mc\ while no bar is present in \md.  We refer readers to \citet{Debattista+17} and \citet{portaluri2017} for details of these two simulations.

\begin{table}
 \centering
 \begin{tabular}{ll}
  \hline
  \textbf{Simulation Name} & \textbf{Details}\\
  \hline
  
  \ma{}& Clumps \& bar \\[2pt]  
  
  \mb{} & Clumps \& no bar \\[2pt]
  
  \mc{} & No clumps \& bar\\[2pt]

  \md{} & No clumps \& decaying bar \\[2pt]  
  \hline
 \end{tabular}
 \caption{The naming convention used reflects the absence/presence of clumps and bars. Whether or not the bar is long lived or decays is also reflected in the name.}
 \label{tab:anysymbols}
\end{table}

\begin{figure*}
	\includegraphics[width=\linewidth]{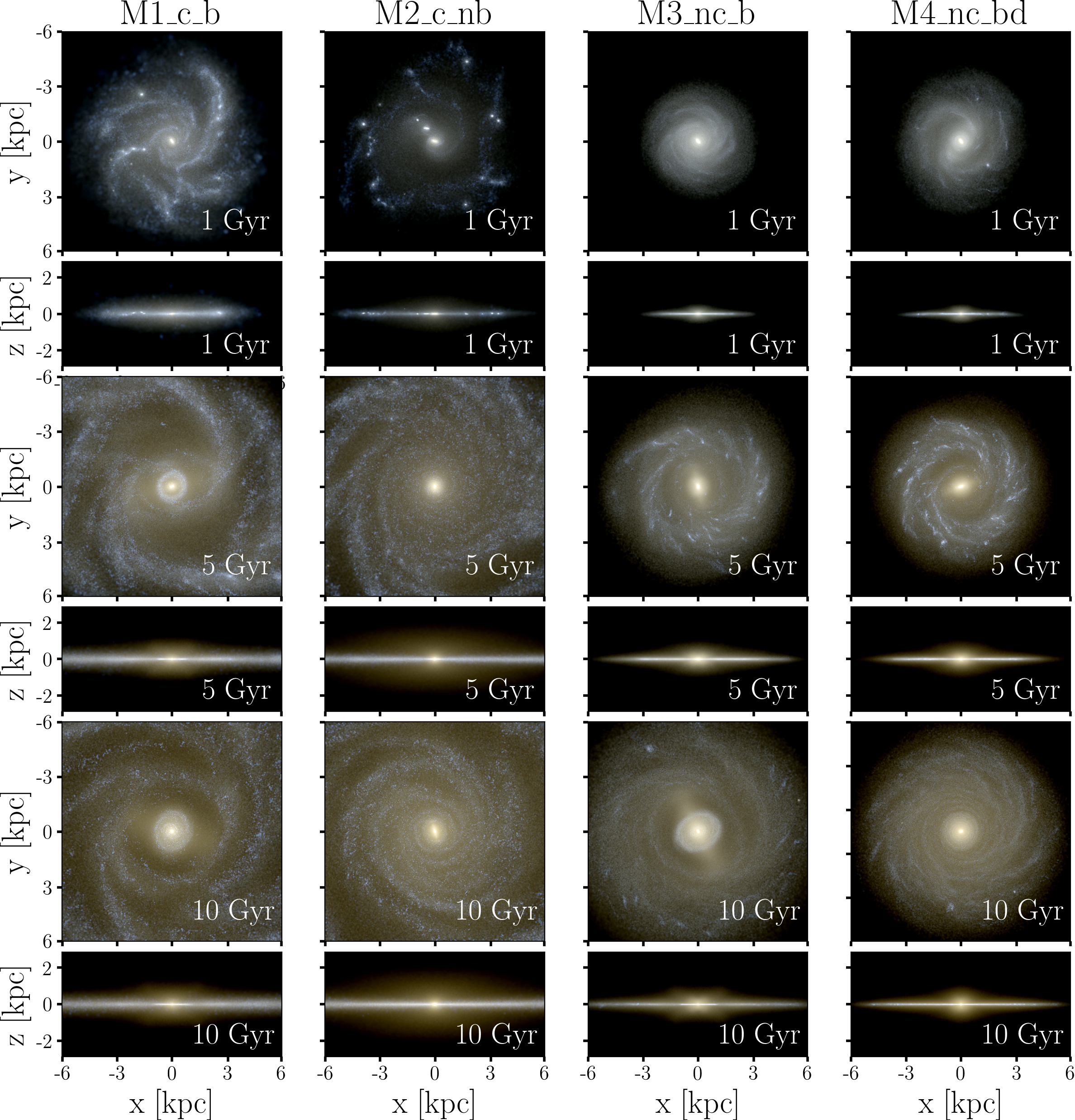}
    \caption{Synthetic RGB images produced with \textsc{pynbody} of the stellar discs of the models. We show the models both face-on and side-on at $1 \Gyr$ (top two rows), then at $5\Gyr$ (next two rows), and at $10\Gyr$ (bottom two rows). These images highlight the different evolutionary paths the models take. The clumpy nature of {\ma} and {\mb} early in their evolution may also be contrasted with {\mc} and {\md}, which do not go through a clumpy episode.}
    \label{fig:sim_renders}
\end{figure*}

\subsection{Simulation details for {\ma} and {\mb}}

The initial conditions of these two models are identical. Their dark matter halo has a virial radius $r_{200} \simeq 200 \kpc$ and a virial mass $M_{200} = 10^{12} \Msun$. The gas corona follows the same radial profile but constitutes only $10\%$ of the mass.  No other baryons (and therefore no stars) are present at $t=0$.  Gas particles are given a tangential velocity with cylindrical rotation such that the spin parameter $\lambda = 0.065$. Here, the spin parameter is defined as $\lambda \equiv J |E|^{1/2}/(GM_{\mathrm{vir}}^{5/2})$, where $J$ and $E$ are the total angular momentum and the energy of the gas particles, and $G$ is the gravitational constant \citep[e.g.][]{peebles1969}. The gas corona and the dark matter halo are comprised of $5 \times 10^6$ particles each, with softening parameters of 
$\epsilon = 50 \pc$ (gas) and $\epsilon = 100 \pc$ (dark matter).  
Star particles forming from the cooling gas also have a softening parameter of $\epsilon = 50 \pc$.  

The two models are evolved using these initial conditions for $10 \Gyr$ with {\sc gasoline} \citep{Wadsleyetal2004, Wadsley2017} using different sub-grid models for the physics of gas cooling, star formation, and supernova feedback. Both {\ma} and {\mb} undergo an episode of clump formation early in their evolution. In the case of {\mb}, the inclusion of metal-line cooling of \citet{ShenWadsleyStinson10} allows the gas to cool more efficiently, which, together with the lower feedback, results in a more vigorous clumpy episode than in model {\ma}. As gas cools, it settles into a disc and once the gas density exceeds 0.1 cm$^{-3}$ star formation commences from gas particles with temperature below 15,000~K which are part of a converging flow.  In both models, star formation efficiency is set to $5\%$. We use the blast wave supernova feedback recipe of \citep{stinson2006}. Supernova feedback couples $40\%$ of the $10^{51}$ erg per supernova to the interstellar medium as thermal energy in {\ma}, and only $10\%$ in {\mb}. Gas mixing uses turbulent diffusion as described by \citet{ShenWadsleyStinson10}. 
We use a base timestep of $\Delta t=5\Myr$ with timesteps refined such that $\delta t = \Delta t/2^n < \eta\sqrt{\epsilon/a_g}$, where we set the refinement parameter $\eta = 0.175$. We set the opening angle of the tree-code gravity calculation to $\theta = 0.7$. Gas particle timesteps also satisfy the condition $\delta t_{gas} = \eta_{\rm{courant}}h/[(1 + \alpha)c + \beta \mu_{\rm{max}}]$, where $\eta_{\rm{courant}} = 0.4$, $h$ is the SPH smoothing length set over the nearest 32 particles, $\alpha$ and $\beta$ are the linear and quadratic viscosity coefficients and $\mu_{\rm{max}}$ is described in \citet{Wadsleyetal2004}.
Note that model {\mb} is comparable to a higher mass resolution version of the clump forming simulation described in \citet{Clarke+19}, \citet{BeraldoeSilva+20}, and \citet{amarante2020}. The setup and simulation details of model {\mc} and {\md} are described at length in \citet{cole+14}, \citet{Debattista+17} and \citet{portaluri2017}; therefore, we do not repeat those details here.

A visual comparison of all four models can be made in Fig.~\ref{fig:sim_renders}, which shows synthetic RGB images of the stellar disc produced using \textsc{pynbody} \citep{pynbody} for each model at different points in time. In the first row, which represents the face-on stellar disc of each model after $1 \Gyr$, the clumpy nature of {\ma} and {\mb} can be seen clearly, as opposed to {\mc} and {\md}, which do not undergo a clumpy episode. Models {\ma}, {\mc}, and {\md} all undergo bar formation, but only {\ma} and {\mc} manage to keep their bars. While {\mc} never undergoes bar destruction, model {\md} (right column) starts to undergo bar dissolution at $\approx 4\Gyr$. Indeed, while a remnant of the bar is still present at $5 \Gyr$ (third row), this is all but gone by the end of the simulation at $10 \Gyr$ (fifth row). The even-numbered rows in Fig.~\ref{fig:sim_renders} show the evolution of the simulations represented in the side-on view. The clumps can be seen occupying the mid-plane in models {\ma} and {\mb} at $1\Gyr$. Finally, these two have a thicker disc compared to {\mc} and {\md} throughout their evolution due to the scattering generated by the clumps.

Ideally we would also compare results from the models against a control model which does not have either a bar or clumps to estimate the contribution of retrograde motion due to numerical heating. However, to some extent all simulations form at least very weak bars. Despite this, as will be shown in Sections~\ref{cr_drivers} and \ref{cr_spatial}, our results indicate that, in the absence of bars and clumps, the fraction of retrograde stars remains constant. Additionally, we find no evidence that retrograde motion is driven by radial heating due to spirals in the models. We conclude that the retrograde fraction driven by numerical noise must be relatively small.

\section{Results}\label{results}

We now compare the models and investigate the properties of the retrograde populations in them. In Section \ref{sim_param} we quantify the evolution of clumpiness and bar strength in the models. In Section \ref{cr_drivers} we investigate to what extent both perturbations produce retrograde orbits. In Section \ref{cr_spatial} we investigate the spatial extent to which bars and clumps drive retrograde motion. Finally, in Section \ref{age_dist} we analyse the age distributions and some of the orbital characteristics of the retrograde stars. 

\subsection{The clumpiness and bar strength of the models}\label{sim_param}

In order to assess the relative efficiencies of bars and clumps at producing retrograde stars, we first need to quantify the strengths of both perturbations. Models {\ma} and {\mb} undergo an early episode of clump formation. To measure the strength and duration of the clumpy episodes, we start by constructing 2D-histograms of the density of the models at each timestep. Each histogram measures $N_x \times N_y = 400 \times 400$ bins, corresponding to a region $20 \times 20$~kpc in size, and represents the mass distribution of the models seen face-on. We identify the clumps as overdensities in the mass distribution by employing the \textsc{find\_peaks} function in the \textsc{photutils} Python package \citep{photutils_ref}. This detects any local maxima in the mass distribution which are above a (manually-set) density threshold, given by $\zeta = M + (150\sigma)$, where $M$ and $\sigma$ are the median and standard deviation of the mass distribution. Having located the peaks, on the face-on view of the models we construct a circle of radius 150~$\pc$ around each clump to isolate the stars constituting it. While the clumps in the models vary in size, we found that a radius of  150~$\pc$ is sufficient to capture the mass contained within them. Moreover, this yielded an average clump mass of $\sim 10^{8} M_{\odot}$ for both of the clumpy models, which is in agreement with high-resolution observations \citep[e.g.][]{cava2018}. The threshold was optimised to ensure that clump detection minimised contamination from other sources, such as spiral arms. We then determine the total stellar mass contained in all the clumps, and normalise it by the total stellar mass of the model during each time step; this gives us the clump mass fraction, $\chi_{\mathrm{clumps}}$, as a function of time. While this method of detecting clumps also picked up the central mass concentrations (bulges) in the models, these were not included in the calculations of the clumpy mass fraction.

\begin{figure}

\includegraphics[width=\columnwidth]{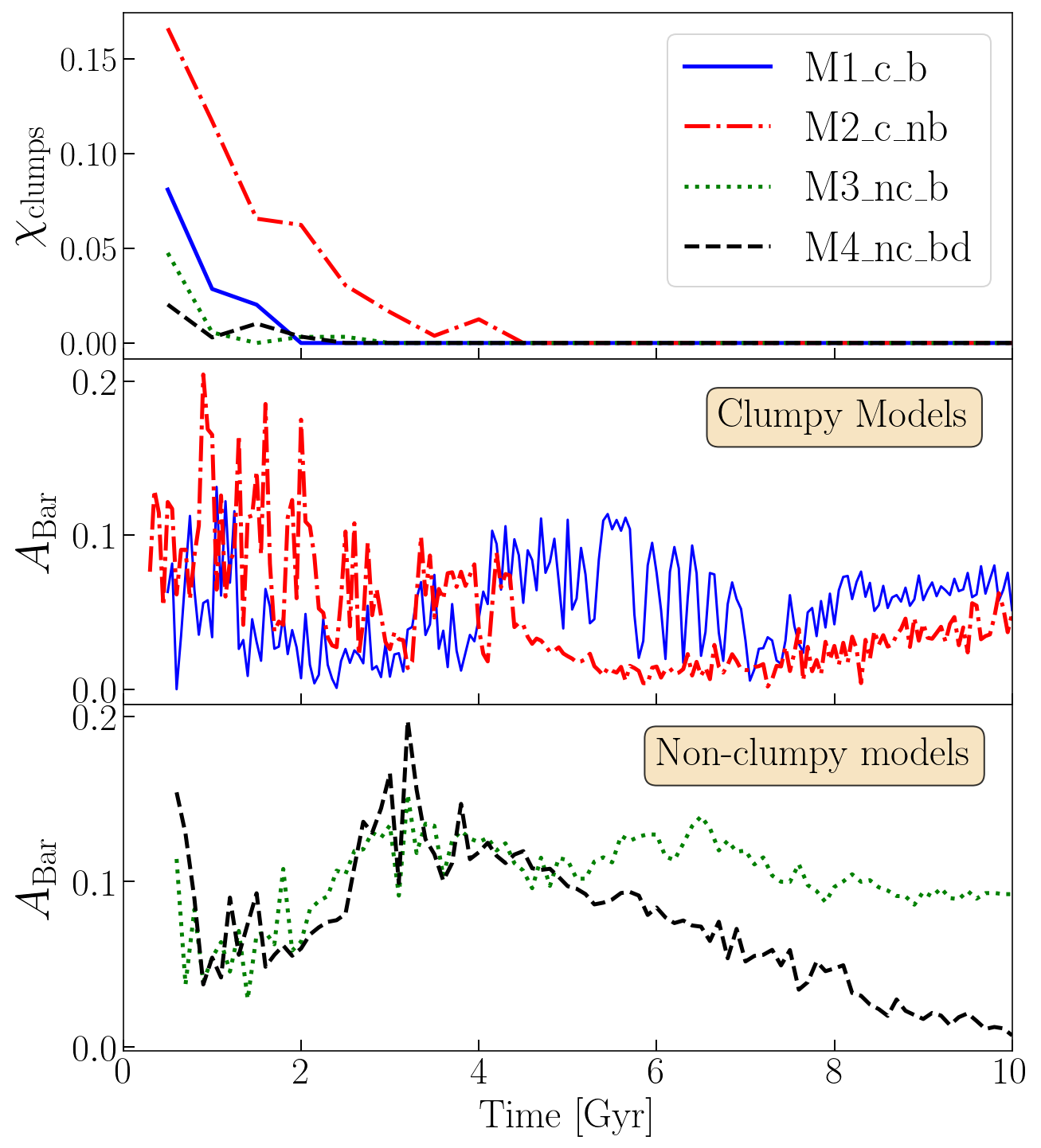}
\caption{Evolution of perturbations in the models. Top: the clumpy mass fraction $\chi_{\mathrm{clumps}}$. While {\ma} and {\mb} both undergo a clumpy episode early in their evolution, M2\_c\_nb has stronger clumping which lasts until about $4.5$~\Gyr. On the other hand, the clumpy episode in {\ma} is over by $2.5$~\Gyr. Models {\mc} and {\md} do not suffer any significant clumping. Middle: evolution of the bar strength, given by the $m = 2$ Fourier moment, for the clumpy models. While bar formation in {\ma} starts at $\approx 4 \Gyr$ and results in a long-lived bar, model {\mb} does not develop a large-scale bar. The non-zero value for $A_{\mathrm{bar}}$ for this model results from the clumps, which show up as $m = 2$ perturbations. Bottom: evolution of the bar strength for the non-clumpy models, {\mc} and {\md}. Both models undergo early bar formation, which peaks at $\approx 4 \Gyr$. However, while the bar in {\mc} is long-lived, the bar in {\md} starts to decay after reaching its peak amplitude, dissolving entirely by the end of simulation.}
\label{fig:spam}
\end{figure}

The top panel of Fig.~\ref{fig:spam} shows the evolution of  $\chi_{\mathrm{clumps}}$ for the four models. Models {\ma} and {\mb} both undergo an early clumpy episode. These clumps can also be seen in the top row of Fig.~\ref{fig:sim_renders}. Overall, model {\mb} is more clumpy than model \ma, and the clumpy episode lasts longer, until $~4.5~\Gyr$ \citep[compare with][]{Clarke+19}, after which no further clumps form. Model {\ma} suffers a milder episode of clump formation, which is over by $2.5 \Gyr$. Models {\mc} and {\md} show no significant clumping throughout their evolution. The non-zero value for $\chi_{\rm{clump}}$ for these models in the first $2\Gyr$ can be attributed largely to spiral contamination, which could not be completely avoided in these models.

The bar strength is measured in the usual way as the amplitude of the global $m=2$ Fourier moment, $A_{\mathrm{Bar}}$, of the face-on mass distribution \citep[e.g.][]{Sellwood1986,  Debattista2006}. The middle and bottom panels of Fig.~\ref{fig:spam} show the evolution of $A_{\mathrm{Bar}}$. Bar formation in {\ma} (middle panel, solid blue line) starts at $\sim 4 \Gyr$ and results in a long-lived bar (see left column of Fig.~\ref{fig:sim_renders}). Model {\mb} (middle panel, dashed red line), on the other hand, does not form a large-scale bar throughout its evolution, as can be seen in Fig.~\ref{fig:sim_renders}. The non-zero value for $A_{\mathrm{Bar}}$ for this model is a result of its clumpy nature in the first few Gyr. Indeed, this signal may be attributed to clumps which happen to be on opposite sides of the disc being picked up as $m=2$ perturbations, resulting in a fake signal. In addition, the bar amplitude for {\mb} also drops as soon as the clumpy episode in {\mb} ends at $4.5 \Gyr$. Spiral arms are also present in models {\ma} and {\mb} (see Fig.~\ref{fig:sim_renders}), which interfere with the $A_{\mathrm{Bar}}$ signal, generating the fluctuations in the data.

In models {\mc} and {\md} (bottom panel of Fig.~\ref{fig:spam}), bar formation starts at $\sim$~2~\Gyr, and reaches peak strength at $\sim$~4~\Gyr. Up until this point in time, the evolution of the bar in both models is very similar. However, the bar in model {\mc} is long-lived, whereas the bar in {\md} slowly starts to weaken at $\sim$~4~\Gyr, eventually dissolving entirely.

\subsection{Bars and clumps as drivers of retrograde motion}\label{cr_drivers}

\begin{figure*}
	\includegraphics[width=\linewidth]{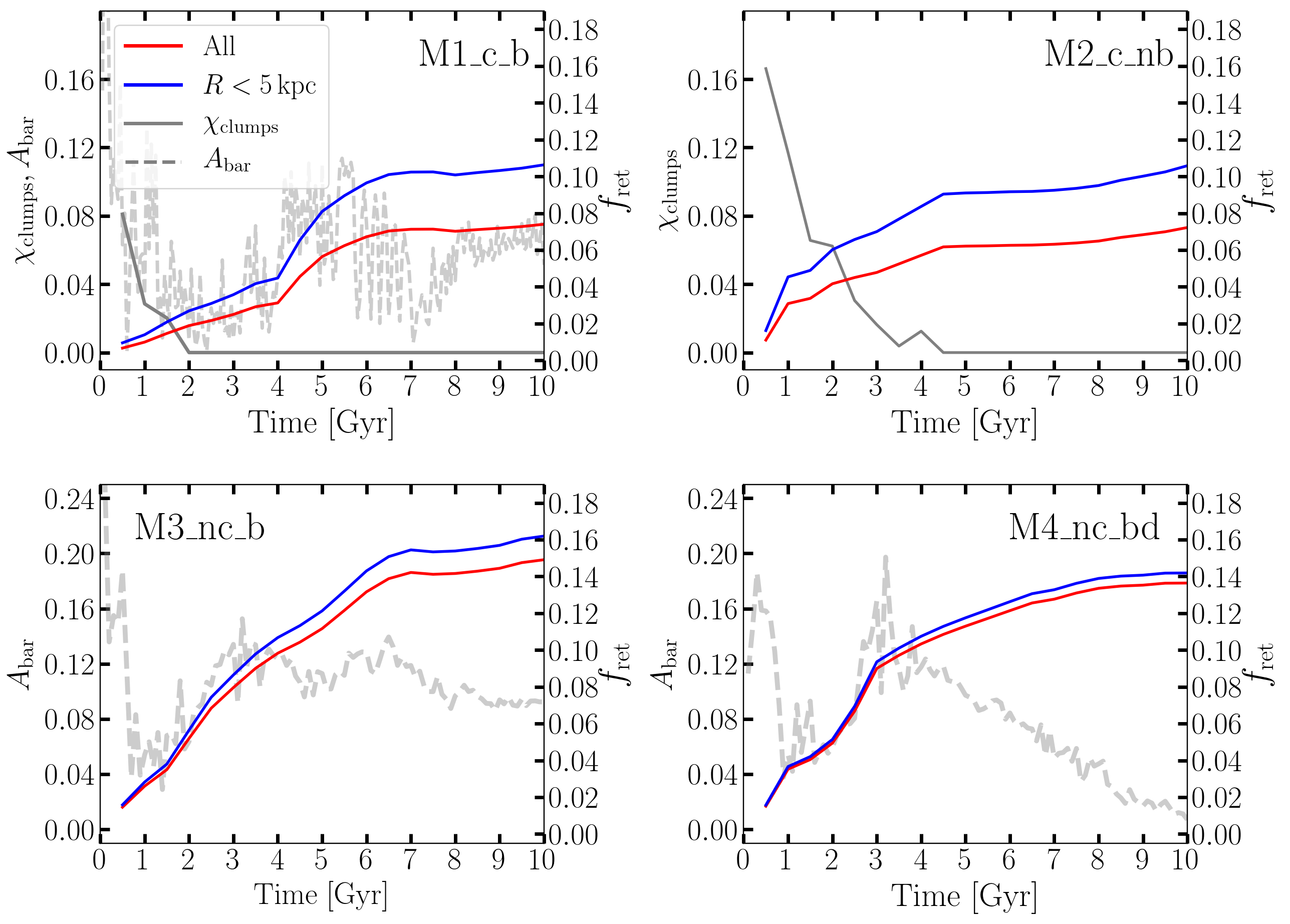}
    \caption{The time evolution of the retrograde mass fraction, $f_{\mathrm{ret}}=M_{\mathrm{ret}}(t)/M_{\star}(\mathrm{10\Gyr})$, in the four models, where $M_{\mathrm{ret}}(t)$ and $M_{\star}(\mathrm{10\Gyr})$ are the total mass of retrograde stars at time $t$ and the total stellar mass at the end of the models, respectively. The red lines show $f_{\mathrm{ret}}$ for all stars, while the blue lines reflect $f_{\mathrm{ret}}$ inside $5\kpc$, with $M_{\star}(\mathrm{10\Gyr})$ being the total stellar mass also inside this region. The solid and dashed grey lines reflect the clumpy mass fraction and the bar amplitude respectively. In both cases the scales for the grey lines are presented on the left hand axis.}
    \label{fig:mfac}
\end{figure*}

In this subsection, we show that bars and stellar clumps play a role in driving retrograde motion of disc stars in the models. Fig.~\ref{fig:mfac} shows the time evolution of the total mass fraction of retrograde stars, $f_{\mathrm{ret}}=M_{\mathrm{ret}}(t)/M_{\star}(\mathrm{10\Gyr})$ in the models, where $M_{\mathrm{ret}}(t)$ is the total mass of retrograde stars at time $t$, and $M_{\star}(\mathrm{10\Gyr})$ is the total stellar mass at the end of the model's evolution ($10\Gyr$), respectively. While the red lines track $f_{\mathrm{ret}}$ for the entire models, the blue lines represent the retrograde mass fraction inside a cylindrical galactocentric radius of $5\kpc$, and are normalised by the total stellar mass in this region. The solid and dashed gray lines reflect the clumpy mass fraction and the bar amplitude respectively.

Model {\mb} (top right panel in Fig.~\ref{fig:mfac}), which forms clumps but no large-scale bar, has a population of retrograde stars which increases continuously up until $4.5 \Gyr$ (red line), when it reaches a plateau. This coincides with the point at which the last clumps in this model dissolve. When considering the entire model (red line), model \mb\ reaches $f_{\mathrm{ret}} \simeq 0.06$ by the end of the clumpy epoch ($4.5\Gyr$). Retrograde stars at $4.5\Gyr$ represent $10\%$ of the total stellar mass at that time. After the clumpy epoch, $f_{\mathrm{ret}}$ remains roughly constant up until $8\Gyr$. This indicates that, in the absence of both clumps and a bar, additional retrograde stars are not being produced in any significant amount via other forms of scattering, whether physical or numerical. The slight increase in the retrograde fraction after $8\Gyr$ can be attributed to the formation of a small, $1\kpc$-scale, bar during this time, which can be seen in Fig.~\ref{fig:sim_renders} (second column). Clumps also drive retrograde motion in {\ma} (top left panel in Fig.~\ref{fig:mfac}). Due to its weaker clump formation episode (grey line), it only reaches $f_{\mathrm{ret}} \simeq 0.03$ during its initial clumpy stage, which ends at roughly $2.5\Gyr$. Retrograde stars at $2.5\Gyr$ represent $6\%$ of the total stellar mass at that time. However, once the bar starts to form at around $4\Gyr$, $f_{\mathrm{ret}}$ roughly doubles during a short time interval ($\sim 2\Gyr$) before reaching a peak, and increases slowly thereafter. 

Models {\mc} and {\md} (bottom row), which do not suffer any significant clumping but form bars early in their evolution, both develop a large retrograde population, with both models reaching $f_{\mathrm{ret}} \simeq 0.14$ by $10\Gyr$. In model {\mc} (bottom left panel),  $f_{\mathrm{ret}}$ increases continuously as the bar grows until roughly $7\Gyr$, when the growth of $f_{\mathrm{ret}}$ slows down, corresponding to the weakening of the bar at this time (see dashed grey line). In model {\md} (bottom right panel), $f_{\mathrm{ret}}$ increases steadily as the bar grows until roughly $3\Gyr$. However, as the bar dissolves (see dashed grey line) we see that while retrograde stars are still being produced, this happens at a continuously declining rate.

The solid blue lines, which reflect $f_{\mathrm{ret}}$ inside $R = 5\kpc$, are normalised by the total stellar mass inside this region. In general, we find that $f_{\mathrm{ret}}$ is higher inside $R = 5\kpc$, indicating that these retrograde populations are centrally concentrated. 

\begin{figure*}
	\includegraphics[width=\linewidth]{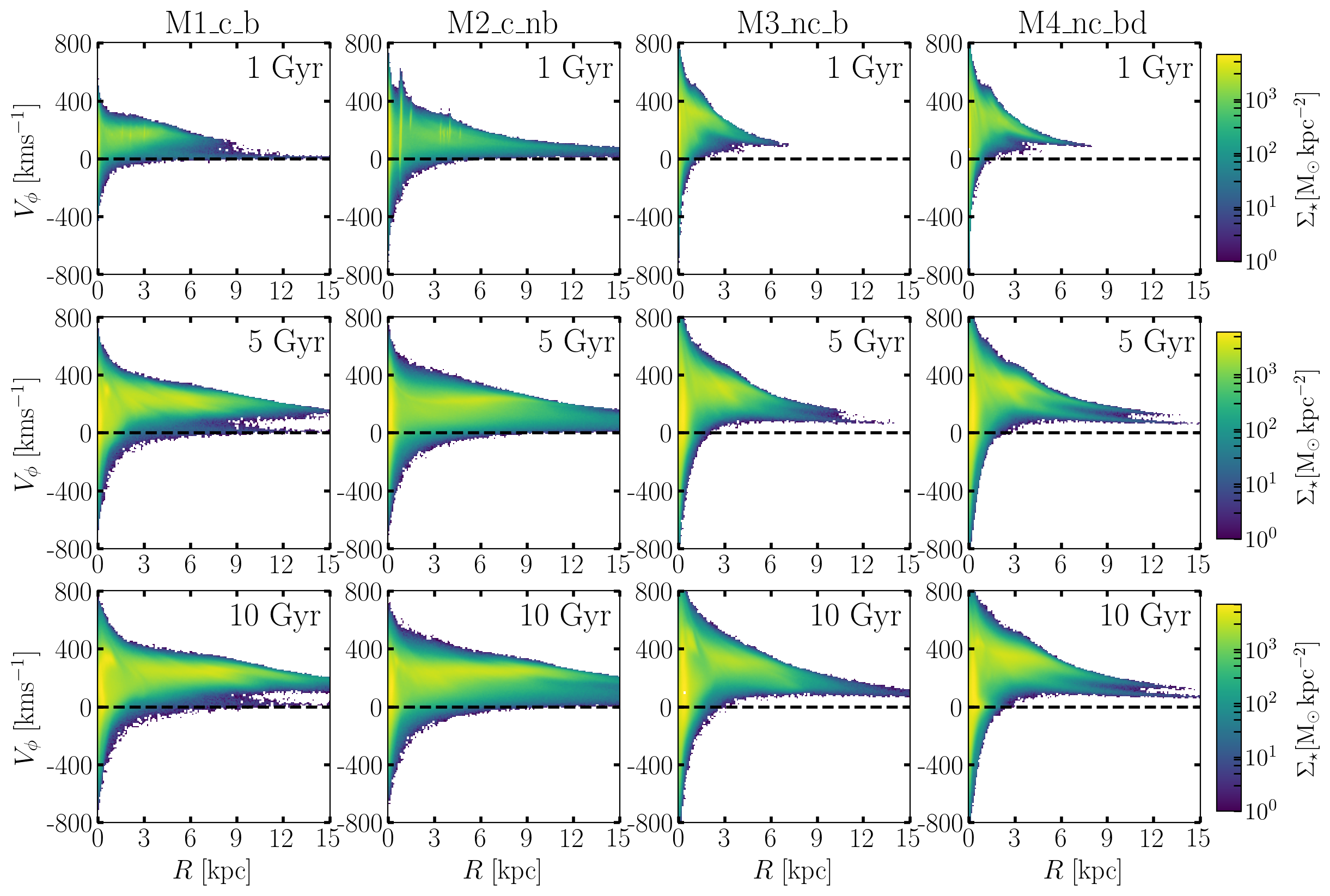}
    \caption{Density distributions in the $V_{\phi}$ versus cylindrical galactocentric radius, $R$, plane, of stars in the models at three different times. The retrograde population in the clumpy model {\ma} (first column) shows itself as a distinct peak centred around $V_{\phi} \approx 0$ and is present even at $1\Gyr$ due to clumps already having scattered stars by this time. This is also the case in model {\mb}, which has stronger clumps. However, the peak centred at $V_{\phi} \approx 0$ is continuous with the main stellar population. The vertical overdensities in {\ma} and {\mb} at $1\Gyr$ are due to the presence of clumps at that time. The non-clumpy models, {\mc} and {\md}, lack a peak at $V_{\phi} \approx 0$, with the retrograde populations being present only in the central regions.}
    \label{fig:vphi_r}
\end{figure*}

Fig.~\ref{fig:vphi_r} shows the distribution of $V_{\phi}$ versus cylindrical galactocentric radius, $R$, for all stars in the models at three different times. Both of the clumpy models, {\ma} and {\mb}, have peaks at $V_{\phi} \approx 0$, even during the earliest time at $1\Gyr$ (first row). This is due to the clumps having already scattered stars by this time. Additionally, while the $V_{\phi} \approx 0$ peak in model {\ma} seems to be distinct from the main stellar population, especially at later times, this is not the case in model {\mb}, which shows a more continuous transition. This is likely due to the stronger clumpy episode in {\mb}, which has driven larger amounts of scattering, and as a result produces a continuous distribution. The vertical overdensities in models {\ma} and {\mb} at $1\Gyr$ are due to the presence of the clumps. On the other hand, the non-clumpy models {\mc} and {\md} are not peaked at  $V_{\phi} \approx 0$, with the retrograde populations being continuous with the main stellar population.

We conclude that both bars and clumps can produce a retrograde population of order $10\%$ (when considering the whole galaxy). This is a fairly significant population. We find that bars are slightly more efficient at producing retrograde stars, but the difference is not very large.

\subsection{The spatial distribution of retrograde stars}\label{cr_spatial}

Having established that both clumps and bars drive stars to retrograde motion, we shift our attention to the spatial distribution of these retrograde stars.  Fig.~\ref{fig:stellardens} shows the radial density profiles, ($\Sigma_{*}$), for the models at different times. The left-hand column represents the density profiles for all stars in the models, while the middle column shows the radial profiles of just the retrograde stars. The left-hand column shows the usual inside-out formation of discs \citep[e.g.][]{chiappini2001,munoz2007, bovy2012, frankel2019}, with the density profile extending to increasingly large radii with the passage of time. This is in part due to the increasingly large gas disc, but is also the result of angular momentum exchange, which leads to the migration of stars to ever larger radii \citep[e.g.][]{roskar2008, sharma2020}.

The middle column of Fig.~\ref{fig:stellardens} shows the differences amongst the four models in the distribution and evolution of the retrograde stars. Firstly, in all cases, the radial density profile of retrograde stars shows only marginal evolution as time progresses when compared to the rest of the stellar populations. At $R \ga 5 \kpc$, the density of retrograde stars at $2\Gyr$ (green line) is higher than at all other times; this reflects on an ongoing spray of stars that has not yet settled into a stationary distribution.

The radial extent of retrograde stars is very different between clumpy and non-clumpy models. This can can be seen in the right column of  Fig.~\ref{fig:stellardens}, where we show the mass fraction of retrograde stars, $f_{\mathrm{ret}}$, in each radial bin, at $10\Gyr$ for each model. In models {\mc} and {\md}, neither of which formed clumps, the retrograde population remains confined within the immediate vicinity of the bar. We also find that bar-driven retrograde stars remain in the region of the bar even after the bar dissolves. On the other hand, the profiles for models {\ma} and {\mb} are more radially extended. The retrograde population is most radially extended in model {\mb}, which had the most vigorous clump-formation episode. However, even the weaker clumpy episode of {\ma} still produces an extended profile, showing that even mild clump formation can cause retrograde stars to be scattered to large radii.

This contrast in the radial extent of bar-driven and clump-driven retrograde stars can also be seen in Fig.~\ref{fig:xzdist}, where we plot the density distribution of these stars in the $(x,z)$ Cartesian plane, for different bins of angular momentum, $L_{\mathrm{z}}$. We see again that the retrograde stars in the clumpy models (first two columns in Fig.~\ref{fig:xzdist}) reach larger galactocentric radii than those in the non-clumpy models (second two columns), for all bins of $L_{\mathrm{z}}$ (i.e.~in all rows). Clump-driven retrograde stars also reach larger heights above the mid-plane, as scattering by clumps converts in-plane motions to vertical ones. In addition, in the bin with the most negative angular momentum (bottom row), the retrograde stars form a relatively flattened distribution, becoming more spherical as $L_z$ increases (middle and top rows).

The lack of retrograde stars in the outer regions ($R > 5\kpc$) of the non-clumpy models rules out the possibility for retrograde motion being driven by radial heating due to spirals. If this were the case, then we would observe a population of retrograde stars in the outer discs of these models where the spirals are present. However, we find no evidence for this.

We conclude that while bars and clumps both drive retrograde motion in disc stars, bars are not efficient at producing retrograde stars at large radii. This tendency for bar-driven retrograde stars to remain in the vicinity of the bar might possibly be due to the presence of the bar resonances. Indeed, stars whose orbits are in the vicinity of the resonances of a growing bar perturbation may be captured into librating orbits \citep[e.g.][]{Kalnajs1973, Tremaine1984, collett1997}. Consequently, the bar-driven retrograde stars will be unable to escape the bar. On the other hand, while a large proportion of the clump-driven retrograde stars also remains in the inner regions of the simulated galaxies, scattering by the clumps is more likely to drive these stars to larger radii.

\subsection{Ages and orbits of retrograde populations}\label{age_dist}

\begin{figure*}
\includegraphics[width=\linewidth]{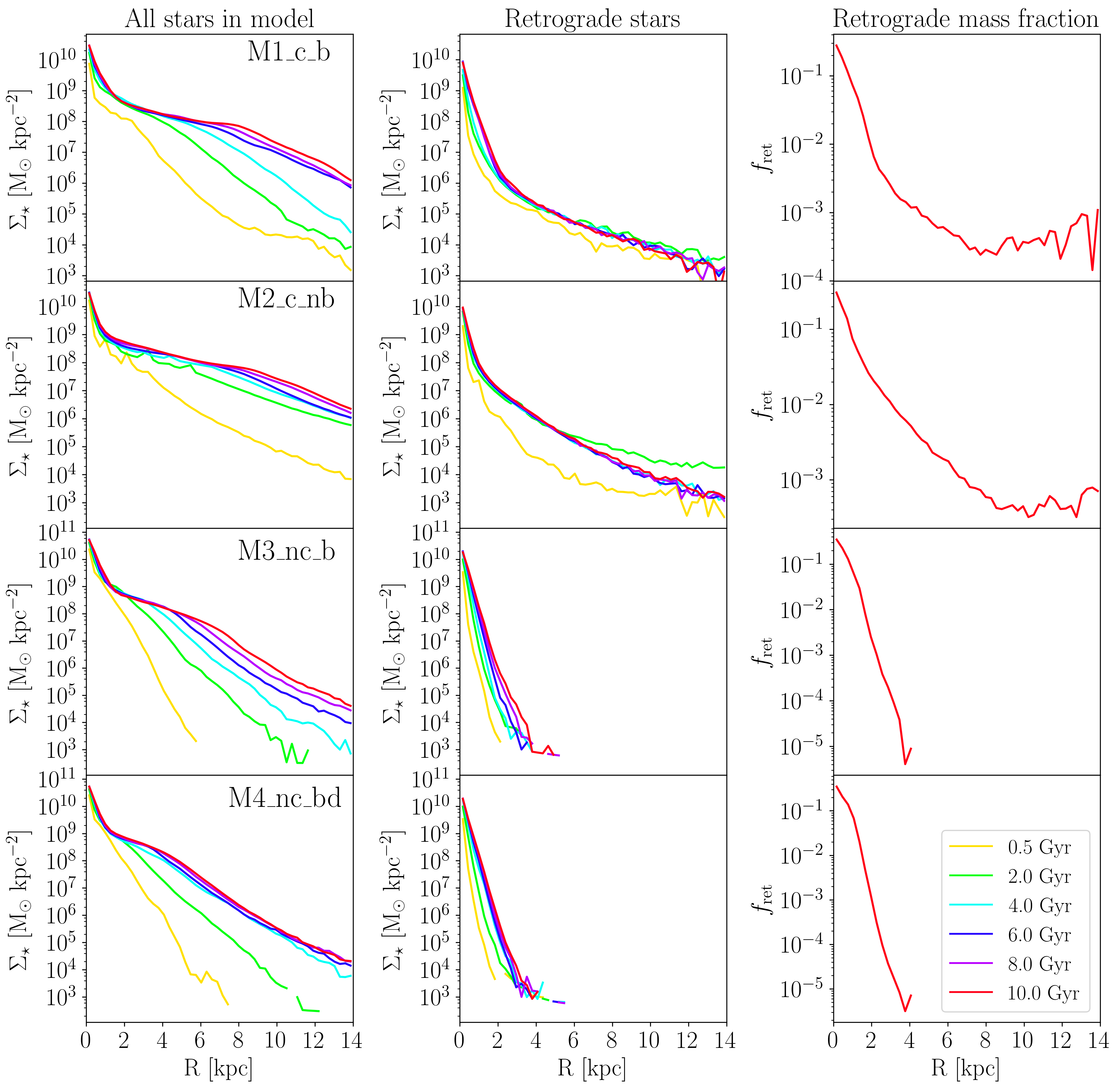}
\caption{Left: the evolution of the radial density profiles for all stars in each model. Middle: the evolution of the radial density profiles for retrograde stars only. Right: the profile of the mass fraction of retrograde stars, $f_{\mathrm{ret}}$, at $10\Gyr$ for each model, which reflects the radial distribution of retrograde stars at the end of each model's evolution. The full profiles show substantial radial growth with time, while the profiles for the retrograde stars exhibit only minimal outward growth. The radial extent of the retrograde stars is strongly dependent on whether clumps were present in the early epochs or not. Even a mild episode of clump formation is sufficient to extend the retrograde stars to larger radii, whereas the retrograde stars driven by the bar tend to remain within the vicinity of the bar.}
    \label{fig:stellardens}
\end{figure*}

Fig.~\ref{fig: tromba} shows the distribution of stellar ages versus the circularity parameter (at $10\Gyr$), defined as  $\lambda_{z} \equiv L_z/L_c(\mathrm{E})$, where $L_c(\mathrm{E})$ is the angular momentum of a particle with binding energy $E$ to be on a circular orbit \citep{abadi+2003}. In computing $L_c(\mathrm{E})$, we assume an axisymmetric disc, and that the orbit lies in the mid-plane. Therefore, some caution must be exercised in interpreting the circularity distribution for stars in the region of the bar. The top row, which presents the distribution for stars in the outer regions ($R > 5\kpc$), shows that in models {\ma} and {\mb}, the retrograde stars were all born during the clumpy epoch (indicated by the horizontal, shaded region). Fig.~\ref{fig:stellardens} showed that bars are unable to drive significant retrograde orbits much beyond their radius. This is also reflected in the upper panels for models {\mc} and {\md} (right columns), where virtually no retrograde stars inhabit this outer region. The bottom row of Fig.~\ref{fig: tromba} shows that retrograde stars in the inner regions ($R < 5\kpc $) span a much wider range of ages. This is expected since a long-lived bar will continuously produce retrograde orbits throughout its life.

\begin{figure*}
\includegraphics[width=\linewidth]{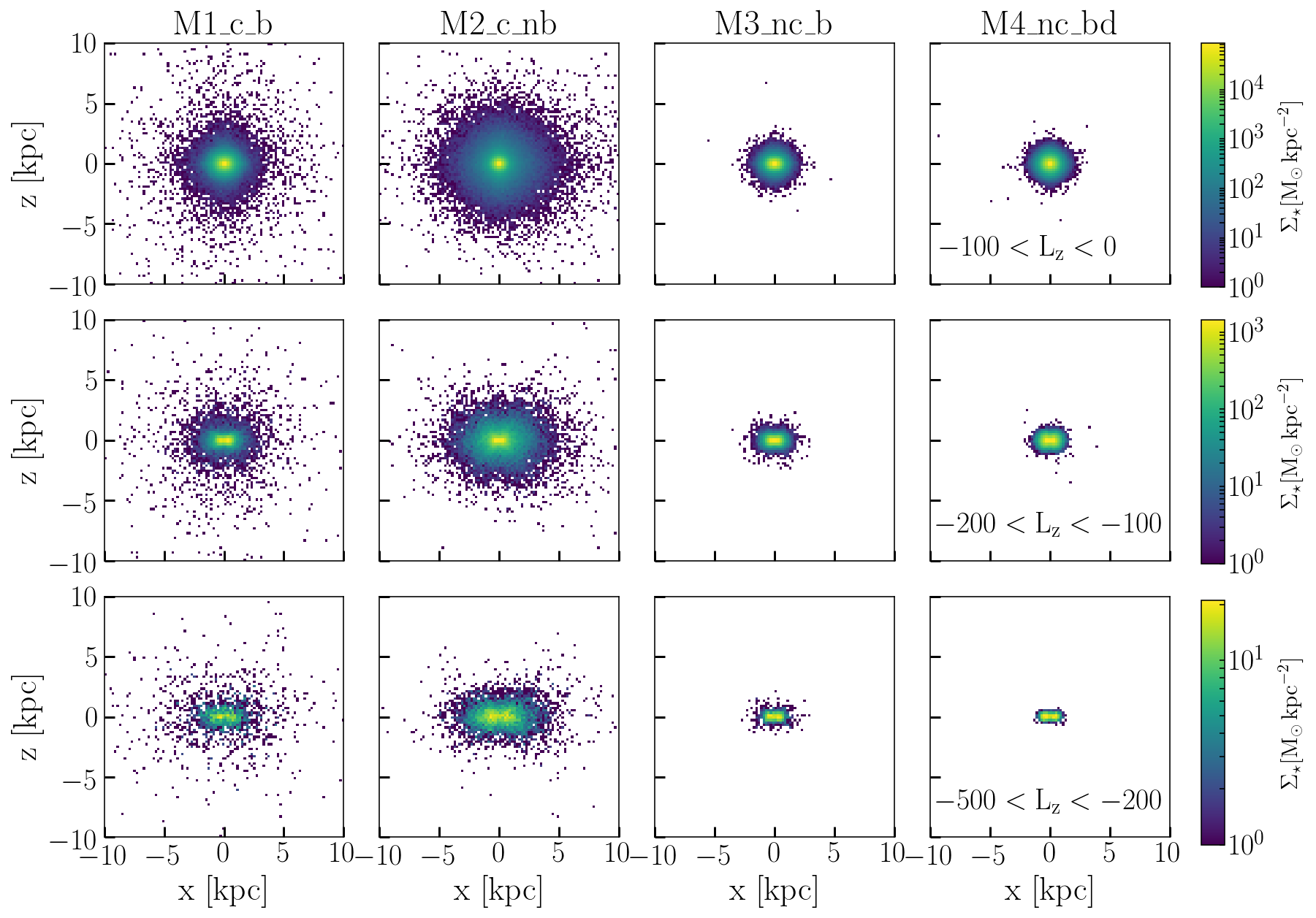}
\caption{Density distribution plots in $(x,z)$ plane of retrograde stars at $10 \Gyr$ for each of the models. Each row represents the density distribution of retrograde stars in different angular momentum ($L_z$) bins. The retrograde stars in {\ma} and {\mb} (first two columns) are found at larger radii due to the clump scattering for all $L_z$ ranges, compared to centrally concentrated retrograde population in {\mc} and {\md}. In addition, it is evident that bars also cause some vertical heating, causing some of the bar-driven retrograde populations to be nearly-spherically distributed. }
    \label{fig:xzdist}
\end{figure*}

\begin{figure*}
\includegraphics[width=\linewidth]{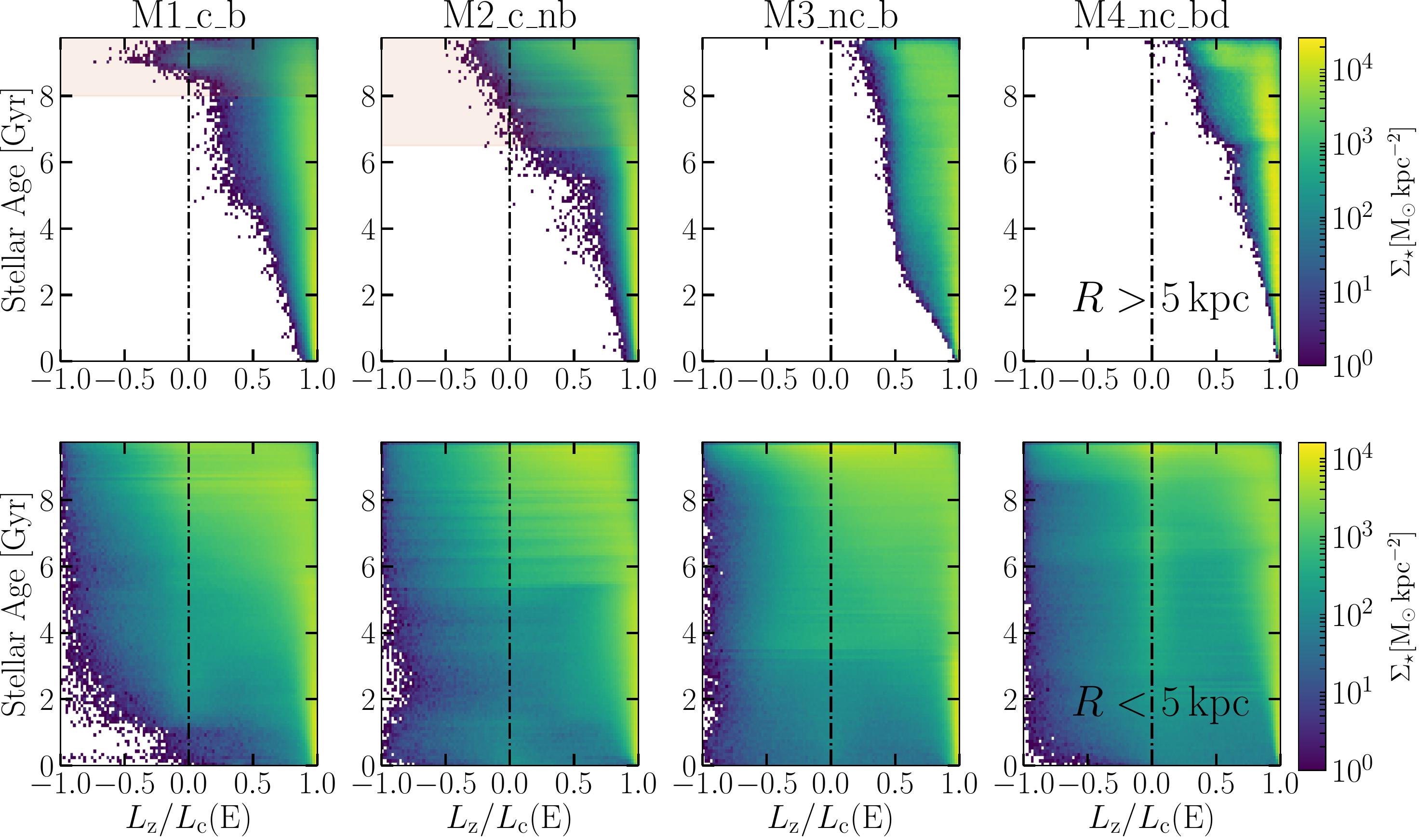}
\caption{Distribution of stellar age versus orbital circularity, $\lambda_z = L_{z}/L_\mathrm{c}(\mathrm{E})$, for stars at $R > 5\kpc$ (top row) and $R < 5\kpc$ (bottom row) at $10\Gyr$. Stars in the shaded regions in the upper row are born during the clumpy episodes of {\ma} and {\mb}. In the outer region, the only retrograde stars present are those which were born during the clumpy epoch.  Models {\mc} and {\md} have no retrograde stars in this outer region. In the inner regions (bottom row), retrograde stars span a much wider range of ages. This is due to the heating of stars by the bar throughout its lifetime. In the case of {\mb}, the presence of younger retrograde stars in the inner region is likely due to a small bar which forms at roughly $8\Gyr$.}
    \label{fig: tromba}
\end{figure*}

\begin{figure*}
	\includegraphics[width=\linewidth]{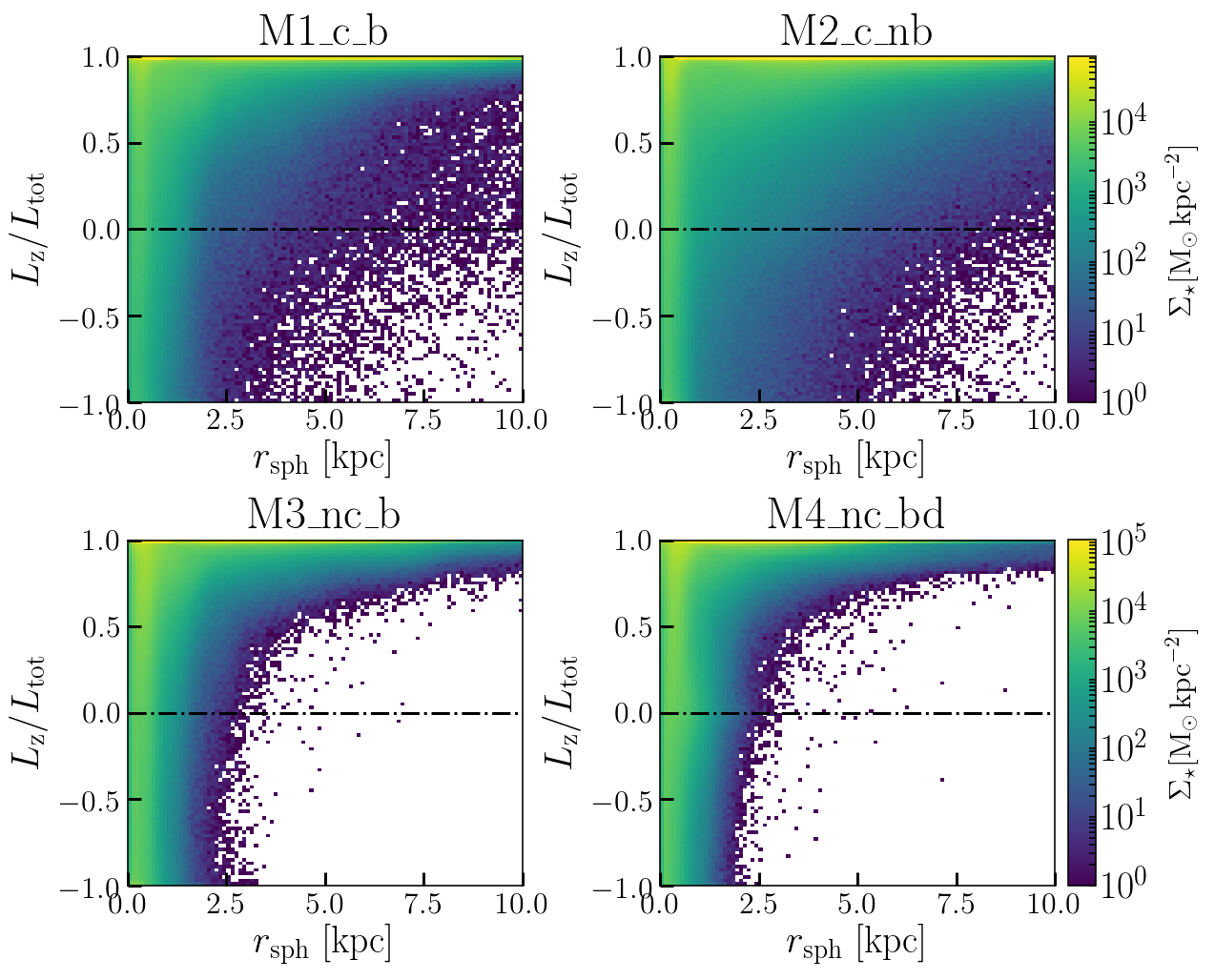}
    \caption{Density distribution in the space of $L_{\mathrm{z}}/L_{\mathrm{tot}}$ versus spherical radius $r_{\mathrm{sph}}$ at $10\Gyr$ for the models. Stars with $L_{\mathrm{z}}/L_{\mathrm{tot}}$ close to unity are on in-plane orbits. On the other hand, stars with  values of  $L_{\mathrm{z}}/L_{\mathrm{tot}}$ close to zero are on vertically heated orbits. The upper panels for the clumpy models show that the majority of retrograde stars in the outer regions (and to a lesser extent the inner regions) are on such vertically heated orbits.  }
    \label{fig:lzltot}
\end{figure*}

Fig.~\ref{fig: tromba} also shows that most of the retrograde orbits produced by clumps in the outer regions have near-zero or low circularities.  This is most apparent in the upper panel for models {\ma} and {\mb}, where retrograde stars have circularity in the range $\lambda_{\mathrm{z}} = -0.5 - 0$. Retrograde stars in the inner regions (bottom row) also span a wider range of $\lambda_{z}$, with the most circular orbits (higher negative $\lambda_{z}$) being attained by the oldest stars. We further characterise the motion of the retrograde populations in Fig.~\ref{fig:lzltot}, where we plot the distribution of retrograde stars in the space of $L_{\mathrm{z}}/L_{\mathrm{tot}}$ versus the spherical radius $r_{\mathrm{sph}}$ for the models at $10\Gyr$. In models {\ma} and {\mb} (upper row), the outer regions ($r_{sph}>5\kpc$) are clearly dominated by retrograde stars having low values of $L_{\mathrm{z}}/L_{\mathrm{tot}}$, which indicates that much of the angular momentum of these stars is not perpendicular to the disc plane. Combined with their low circularity, it suggests that clump-driven retrograde stars are predominantly on boxy orbits. On the other hand, orbits span a wider range of  $L_{\mathrm{z}}/L_{\mathrm{tot}}$ in the inner region where the bar dominates. However, we have also seen in Fig.~\ref{fig:xzdist} that bars produce a retrograde population which has a spheroidal, rather than a discy, distribution.

\section{Observational consequences}\label{conc}

We now consider the observational implications of our results for the Milky Way, and for disc galaxies in general.

\subsection{Implications for the Milky Way} \label{sec:mw}

We have demonstrated that while bar-driven retrograde stars remain in the vicinity of the bar, clumps are capable of driving retrograde motion to larger galactocentric radii (Figs.~\ref{fig:stellardens},~\ref{fig:xzdist}). Therefore, we expect that a population of clump-driven retrograde stars will be present in the Solar neighbourhood if the Milky Way has experienced a clumpy episode. On the other hand, we do not expect any bar-driven retrograde stars to be present in this region, as these will be restricted to galactocentric radii comparable to the Milky Way's bar. In the clumpy models we found the retrograde mass fraction, $f_{ret}$, in the Solar cylinder (defined as $7.5<R_{G}/{\rm kpc}<8.5$ and $0< |z|/{\rm kpc} <2$) to be $0.01\%$ and $0.04\%$ for {\ma} and {\mb} respectively. The clump-driven retrograde stars in the models are part of the Splash-like population studied in \citet{amarante2020}. They found the relative fraction of this population, defined as the metal-rich (-0.7$<$[Fe/H]$<$-0.2), low angular momentum ($v_{\phi}<100\kms$) stars in an isolated clumpy simulation\footnote{\citet{amarante2020} studied a lower mass resolution version of {\mb}.} is in good agreement with the fraction observed in the Milky Way using the same selection criteria (see their figure 4, left panel). 

We estimate the Milky Way's Solar neighbourhood retrograde fraction (those within $1\kpc$ of the Sun) based on the local fraction of the kinematically-defined halo, which is $0.45-0.60\%$ \citep{kordopatis2013, posti2018, amarante2020a}. Since this population's velocity distribution is well fit by a Gaussian, with mean $v_{\phi} \sim 20\kms$ and $\sigma_\phi = 85-100 \kms$, roughly $40\%$ of these stars are on retrograde orbits. The resulting fraction of retrograde stars in the Solar neighbourhood therefore is $\approx 0.18-0.24\%$. This fraction has no chemical selection; for this reason, its higher value compared to the models is attributed to the fact that it also includes halo stars accreted by the Milky Way (e.g. \citealt{hawkins2015, hayes2018, helmi2018, belokurov2018, Das2020}). The models only have in-situ retrograde stars, and their lower fraction shows that retrograde orbits are not over-produced due to excessive scattering in the models. Moreover, a considerable number of retrograde stars associated with the \textit{Gaia}-Enceladus-Sausage event \citep{belokurov2018, helmi2018} lie exactly in the region associated with the thick disc in the [Mg/Fe]-[Fe/H] plane (see, e.g., \citealt{kordopatis2020} figure 11). Therefore, determining the relative fractions of accreted and in-situ retrograde stars in the Milky Way requires detailed chemical analysis which is beyond the scope of this work.

The inner regions of the Milky Way, which are difficult to study due to extinction and over-crowding, are currently being explored in greater detail. \citet{queiroz2020} used {\it Gaia} DR2, along with chemistry from APOGEE DR16, to explore the bulge/bar region of the Milky Way. They reported a highly eccentric, retrograde component within $1\kpc$ of the Galactic centre. While this inner retrograde component can indeed be generated by clumps sinking into the Galactic centre, the bar, as we have shown, will also play a strong role in building up this population (see Fig.~\ref{fig:mfac}). We further demonstrated in Sec.~\ref{cr_spatial} that bar-driven retrograde stars remain within the bar radius, and will therefore be centrally concentrated (bottom two rows in Figs.~\ref{fig:stellardens} and \ref{fig:xzdist}), and we found in Sec.~\ref{cr_drivers} that bars and clumps contribute roughly equally to the retrograde fraction in the inner regions. 

\subsection{Implications for disc galaxies in general}\label{sec:imp_disc}

We found that the retrograde populations generated by the bar and clumps in the models do not produce the counter-rotating discs observed in disc galaxies such as NGC~4550 \citep{rubin1992, rix1992} and  NGC~7217 \citep{merrifield1994}. Indeed, we have shown in Sec.~\ref{age_dist} that the retrograde stars more closely resemble a heated population in terms of their kinematics, rather than a rotationally-supported disc. This can also be seen in Fig.~\ref{fig:xzdist}, which illustrates how both clump and bar-driven retrograde stars are spheroidally distributed. The kinematically hot orbits combined with the paucity of retrograde stars in the outer disc region make it unlikely that such a retrograde population would have been detected in earlier surveys,  which mainly used line-of-sight velocities to disentangle the counter-rotating populations from the main disc. In one such study, \citet{kuijken1996} observed a sample of 28 S0 galaxies, including (at least) eight barred examples. They found that the galaxies in their sample must have a counter-rotation mass fraction, $f_{cr} \la 0.05$. In addition, they estimated that \la 10\% of S0 galaxies have a significant retrograde population. However, in their study, only spectra in the flat portion of the rotation curves of their galaxies were considered, which means that the bulge/bar region was largely unexplored. Therefore, any retrograde population generated by the bar would remain undetected. In addition, any clump-driven retrograde stars in the outer disc would not be detected either, given how sparse they are in this region (see right column in Fig~\ref{fig:stellardens}). In more recent work, \citet{zhu2018} constructed orbit-superposition Schwarzschild models of galaxies observed in the CALIFA survey \citep{sanchez2012} and classified their orbits according to the circularity, $\lambda_{z}$. In general, they found that more massive galaxies tend to have higher fractions of both hot ($\lvert \lambda_{z} \rvert <0.25$) and counter-rotating ($\lambda_{z}<-0.25$) orbits. However, from their sample, it is unclear whether there is any significant difference in the orbital configuration of barred and unbarred discs. To that end, it would be interesting to see if any future observational studies are able to confirm whether bars do indeed drive retrograde motion in the inner regions of disc galaxies, as the models predict.

Our findings have important implications with regard to the origin of stellar counter-rotating discs. It is widely accepted that this phenomenon is produced by externally driven mechanisms such as the accretion of counter rotating gas which subsequently forms stars \citep[e.g][]{cocatto2011,cocatto2013,johnston2013,pizzella2014,pizzella2018}, a scenario which is supported by numerical simulations  \citep{thakar1996,thakar1998,algorry2014}, or possibly from merger events \citep{puerari2001}. However, there have also been efforts to link the phenomenon of counter-rotation with internally driven mechanisms. For example, the `wave patterns' observed in the velocity curves of barred S0 galaxies \citep{bettoni1989, bettoni1997, zeilinger2001} were interpreted by  \citet{wozniak1997} to be the result of a counter-rotating population induced by the bar, and they further suggested that such a population might constitute $14-30\%$ of the stellar mass inside the bar's co-rotation resonance. \citet{evans1994} also showed that it is possible for bar dissolution to generate a counter-rotating population. In this scenario, stars on box orbits are able to escape the potential well to go on tube orbits as the potential becomes axisymmetric. In a non-rotating disc, there are as many stars on co-rotating boxy orbits as there are counter-rotating. Therefore, half of the stars are scattered into counter-rotating tube orbits. We have shown in Sec.~\ref{cr_drivers} that bars are capable of generating a substantial retrograde population, in agreement with \citet{wozniak1997}. However, the bar-driven retrograde stars we find in the models are not counter-rotating, and are restricted to the region of the bar. Furthermore, we showed that bar destruction is unable to produce counter-rotating discs.

\section{Summary}
\label{summ}

By analysing four $N$-body+SPH models of isolated disc galaxies, we investigated the role which stellar clumps and bars play in driving retrograde motion in disc stars. Our results can be summarised as follows:

\begin{enumerate}
    \item Both clumps and bars can generate quite a significant retrograde population (of order $10\%$ of the total stellar mass), with bars being slightly more efficient at producing retrograde stars. We find no evidence that bar dissolution drives additional retrograde motion  (see Sec.~\ref{cr_drivers}).
    
    \item Retrograde orbits generated by a bar remain confined to the inner regions inside the bar radius (even if the bar dissolves). Thus, any stars on retrograde motion found in the Solar neighbourhood cannot be attributed to the bar  (see Sec.~\ref{cr_spatial}).
    
    \item Retrograde orbits generated by clumps reach larger Galactocentric radii, and may be found in the Solar neighbourhood  (see Sec.~\ref{cr_spatial}).
    
    \item The retrograde orbits generated by clumps in the simulations represent $0.01-0.04\%$ of the stars in the mock Solar neighbourhood. This fraction is lower than the estimated in the MW (which includes both accreted and in-situ stars), suggesting that the models do not over-produce retrograde stars (see Sec.~\ref{sec:mw}).
    
    \item Neither bars nor clumps can produce rotationally supported counter-rotating discs. Moreover, we find no evidence that bar destruction produces counter-rotation (see Sec.~\ref{sec:imp_disc}).
    
\end{enumerate}

\section*{Acknowledgements}
The work reported on in this publication has been partially supported by COST Action CA18104: MW-Gaia. K.F. is partially funded by the the Tertiary Education
Scholarships Scheme (TESS, Malta). K.F. thanks the Jeremiah Horrocks Institute for hospitality during a visit.
V.P.D. and L.B.S. are supported by STFC Consolidated grant \#ST/R000786/1. The simulations in this paper were run at the High Performance Computing Facility of the University of Central Lancashire and at the DiRAC Shared Memory Processing system at the University of Cambridge, operated by the COSMOS Project at the Department of Applied Mathematics and Theoretical Physics on behalf of the STFC DiRAC HPC Facility (www.dirac.ac.uk). This equipment was funded by BIS National E-infrastructure capital grant ST/J005673/1, STFC capital grant ST/H008586/1 and STFC DiRAC Operations grant ST/K00333X/1. DiRAC is part of the National E-Infrastructure.
J.A. acknowledges The World Academy of Sciences and the Chinese Academy of Sciences for the CAS-TWAS scholarship.

This research made use of Photutils, an Astropy package for
detection and photometry of astronomical sources \citep{photutils_ref}.

We thank the anonymous referee for their constructive comments which helped to improve this paper.


\section*{Data availability}

The simulation data underlying this article are a mix of proprietary and available; some data may be shared upon reasonable request to V.P.D. (vpdebattista@gmail.com). 



\bibliographystyle{mnras}
\bibliography{refs} 

\begin{thebibliography}{}
\makeatletter
\relax
\def\mn@urlcharsother{\let\do\@makeother \do\$\do\&\do\#\do\^\do\_\do\%\do\~}
\def\mn@doi{\begingroup\mn@urlcharsother \@ifnextchar [ {\mn@doi@}
  {\mn@doi@[]}}
\def\mn@doi@[#1]#2{\def\@tempa{#1}\ifx\@tempa\@empty \href
  {http://dx.doi.org/#2} {doi:#2}\else \href {http://dx.doi.org/#2} {#1}\fi
  \endgroup}
\def\mn@eprint#1#2{\mn@eprint@#1:#2::\@nil}
\def\mn@eprint@arXiv#1{\href {http://arxiv.org/abs/#1} {{\tt arXiv:#1}}}
\def\mn@eprint@dblp#1{\href {http://dblp.uni-trier.de/rec/bibtex/#1.xml}
  {dblp:#1}}
\def\mn@eprint@#1:#2:#3:#4\@nil{\def\@tempa {#1}\def\@tempb {#2}\def\@tempc
  {#3}\ifx \@tempc \@empty \let \@tempc \@tempb \let \@tempb \@tempa \fi \ifx
  \@tempb \@empty \def\@tempb {arXiv}\fi \@ifundefined
  {mn@eprint@\@tempb}{\@tempb:\@tempc}{\expandafter \expandafter \csname
  mn@eprint@\@tempb\endcsname \expandafter{\@tempc}}}

\bibitem[\protect\citeauthoryear{{Abadi}, {Navarro}, {Steinmetz}  \&
  {Eke}}{{Abadi} et~al.}{2003}]{abadi+2003}
{Abadi} M.~G.,  {Navarro} J.~F.,  {Steinmetz} M.,   {Eke} V.~R.,  2003, \mn@doi
  [\apj] {10.1086/378316}, \href
  {https://ui.adsabs.harvard.edu/abs/2003ApJ...597...21A} {597, 21}

\bibitem[\protect\citeauthoryear{{Algorry}, {Navarro}, {Abadi}, {Sales},
  {Steinmetz}  \& {Piontek}}{{Algorry} et~al.}{2014}]{algorry2014}
{Algorry} D.~G.,  {Navarro} J.~F.,  {Abadi} M.~G.,  {Sales} L.~V.,  {Steinmetz}
  M.,   {Piontek} F.,  2014, \mn@doi [\mnras] {10.1093/mnras/stt2154}, \href
  {https://ui.adsabs.harvard.edu/abs/2014MNRAS.437.3596A} {437, 3596}

\bibitem[\protect\citeauthoryear{{Amarante}, {Smith}  \& {Boeche}}{{Amarante}
  et~al.}{2020a}]{amarante2020a}
{Amarante} J. A.~S.,  {Smith} M.~C.,   {Boeche} C.,  2020a, \mn@doi [\mnras]
  {10.1093/mnras/staa077}, \href
  {https://ui.adsabs.harvard.edu/abs/2020MNRAS.492.3816A} {492, 3816}

\bibitem[\protect\citeauthoryear{{Amarante}, {Beraldo e Silva}, {Debattista}
  \& {Smith}}{{Amarante} et~al.}{2020b}]{amarante2020}
{Amarante} J. A.~S.,  {Beraldo e Silva} L.,  {Debattista} V.~P.,   {Smith}
  M.~C.,  2020b, \mn@doi [\apjl] {10.3847/2041-8213/ab78a4}, \href
  {https://ui.adsabs.harvard.edu/abs/2020ApJ...891L..30A} {891, L30}

\bibitem[\protect\citeauthoryear{{Athanassoula}}{{Athanassoula}}{2013}]{athanassoula2013}
{Athanassoula} E.,  2013, {Bars and secular evolution in disk galaxies:
  Theoretical input}.
Cambridge Univ. Press, Cambridge, p.~305

\bibitem[\protect\citeauthoryear{{Athanassoula}, {Dehnen}  \&
  {Lambert}}{{Athanassoula} et~al.}{2005}]{athanassoula2005}
{Athanassoula} E.,  {Dehnen} W.,   {Lambert} J.~C.,  2005, Highlights of
  Astronomy, \href {https://ui.adsabs.harvard.edu/abs/2005HiA....13..343A} {13,
  343}

\bibitem[\protect\citeauthoryear{{Belokurov}, {Erkal}, {Evans}, {Koposov}  \&
  {Deason}}{{Belokurov} et~al.}{2018}]{belokurov2018}
{Belokurov} V.,  {Erkal} D.,  {Evans} N.~W.,  {Koposov} S.~E.,   {Deason}
  A.~J.,  2018, \mn@doi [\mnras] {10.1093/mnras/sty982}, \href
  {https://ui.adsabs.harvard.edu/#abs/2018MNRAS.478..611B} {478, 611}

\bibitem[\protect\citeauthoryear{{Belokurov}, {Sanders}, {Fattahi}, {Smith},
  {Deason}, {Evans}  \& {Grand }}{{Belokurov} et~al.}{2020}]{Belokurov2019}
{Belokurov} V.,  {Sanders} J.~L.,  {Fattahi} A.,  {Smith} M.~C.,  {Deason}
  A.~J.,  {Evans} N.~W.,   {Grand } R. J.~J.,  2020, \mn@doi [\mnras]
  {10.1093/mnras/staa876}, \href
  {https://ui.adsabs.harvard.edu/abs/2020MNRAS.494.3880B} {494, 3880}

\bibitem[\protect\citeauthoryear{{Beraldo e Silva}, {Debattista}, {Nidever},
  {Amarante}  \& {Garver}}{{Beraldo e Silva} et~al.}{2020a}]{BeraldoeSilva+20b}
{Beraldo e Silva} L.,  {Debattista} V.~P.,  {Nidever} D.,  {Amarante} J.,
  {Garver} B.,  2020a, arXiv e-prints, \href
  {https://ui.adsabs.harvard.edu/abs/2020arXiv200903346B} {p. arXiv:2009.03346}

\bibitem[\protect\citeauthoryear{{Beraldo e Silva}, {Debattista},
  {Khachaturyants}  \& {Nidever}}{{Beraldo e Silva}
  et~al.}{2020b}]{BeraldoeSilva+20}
{Beraldo e Silva} L.,  {Debattista} V.~P.,  {Khachaturyants} T.,   {Nidever}
  D.,  2020b, \mn@doi [\mnras] {10.1093/mnras/staa065}, \href
  {https://ui.adsabs.harvard.edu/abs/2020MNRAS.492.4716B} {492, 4716}

\bibitem[\protect\citeauthoryear{{Bettoni}}{{Bettoni}}{1989}]{bettoni1989}
{Bettoni} D.,  1989, \mn@doi [\aj] {10.1086/114958}, \href
  {https://ui.adsabs.harvard.edu/abs/1989AJ.....97...79B} {97, 79}

\bibitem[\protect\citeauthoryear{{Bettoni} \& {Galletta}}{{Bettoni} \&
  {Galletta}}{1997}]{bettoni1997}
{Bettoni} D.,  {Galletta} G.,  1997, \mn@doi [\aaps] {10.1051/aas:1997180},
  \href {https://ui.adsabs.harvard.edu/abs/1997A&AS..124...61B} {124, 61}

\bibitem[\protect\citeauthoryear{{Bournaud}, {Combes}  \& {Semelin}}{{Bournaud}
  et~al.}{2005}]{bournaud2005}
{Bournaud} F.,  {Combes} F.,   {Semelin} B.,  2005, \mn@doi [\mnras]
  {10.1111/j.1745-3933.2005.00096.x}, \href
  {https://ui.adsabs.harvard.edu/abs/2005MNRAS.364L..18B} {364, L18}

\bibitem[\protect\citeauthoryear{{Bournaud}, {Elmegreen}  \&
  {Martig}}{{Bournaud} et~al.}{2009}]{bournaud2009}
{Bournaud} F.,  {Elmegreen} B.~G.,   {Martig} M.,  2009, \mn@doi [\apjl]
  {10.1088/0004-637X/707/1/L1}, \href
  {https://ui.adsabs.harvard.edu/abs/2009ApJ...707L...1B} {707, L1}

\bibitem[\protect\citeauthoryear{{Bovy}, {Rix}, {Liu}, {Hogg}, {Beers}  \&
  {Lee}}{{Bovy} et~al.}{2012}]{bovy2012}
{Bovy} J.,  {Rix} H.-W.,  {Liu} C.,  {Hogg} D.~W.,  {Beers} T.~C.,   {Lee}
  Y.~S.,  2012, \mn@doi [\apj] {10.1088/0004-637X/753/2/148}, \href
  {https://ui.adsabs.harvard.edu/abs/2012ApJ...753..148B} {753, 148}

\bibitem[\protect\citeauthoryear{{Bradley} et~al.,}{{Bradley}
  et~al.}{2020}]{photutils_ref}
{Bradley} L.,  et~al., 2020, {astropy/photutils: 1.0.1},
  \mn@doi{10.5281/zenodo.596036}

\bibitem[\protect\citeauthoryear{{Buta} \& {Combes}}{{Buta} \&
  {Combes}}{1996}]{buta1996}
{Buta} R.,  {Combes} F.,  1996, \fcp, \href
  {https://ui.adsabs.harvard.edu/abs/1996FCPh...17...95B} {17, 95}

\bibitem[\protect\citeauthoryear{{Cava}, {Schaerer}, {Richard},
  {P{\'e}rez-Gonz{\'a}lez}, {Dessauges-Zavadsky}, {Mayer}  \&
  {Tamburello}}{{Cava} et~al.}{2018}]{cava2018}
{Cava} A.,  {Schaerer} D.,  {Richard} J.,  {P{\'e}rez-Gonz{\'a}lez} P.~G.,
  {Dessauges-Zavadsky} M.,  {Mayer} L.,   {Tamburello} V.,  2018, \mn@doi
  [Nature Astronomy] {10.1038/s41550-017-0295-x}, \href
  {https://ui.adsabs.harvard.edu/abs/2018NatAs...2...76C} {2, 76}

\bibitem[\protect\citeauthoryear{{Ceverino}, {Dekel}  \& {Bournaud}}{{Ceverino}
  et~al.}{2010}]{ceverino2010}
{Ceverino} D.,  {Dekel} A.,   {Bournaud} F.,  2010, \mn@doi [\mnras]
  {10.1111/j.1365-2966.2010.16433.x}, \href
  {https://ui.adsabs.harvard.edu/abs/2010MNRAS.404.2151C} {404, 2151}

\bibitem[\protect\citeauthoryear{{Chiappini}, {Matteucci}  \&
  {Romano}}{{Chiappini} et~al.}{2001}]{chiappini2001}
{Chiappini} C.,  {Matteucci} F.,   {Romano} D.,  2001, \mn@doi [\apj]
  {10.1086/321427}, \href
  {https://ui.adsabs.harvard.edu/abs/2001ApJ...554.1044C} {554, 1044}

\bibitem[\protect\citeauthoryear{{Chiba}, {Friske}  \& {Sch{\"o}nrich}}{{Chiba}
  et~al.}{2021}]{Chiba2021}
{Chiba} R.,  {Friske} J. K.~S.,   {Sch{\"o}nrich} R.,  2021, \mn@doi [\mnras]
  {10.1093/mnras/staa3585}, \href
  {https://ui.adsabs.harvard.edu/abs/2021MNRAS.500.4710C} {500, 4710}

\bibitem[\protect\citeauthoryear{{Clarke} et~al.,}{{Clarke}
  et~al.}{2019}]{Clarke+19}
{Clarke} A.~J.,  et~al., 2019, \mn@doi [\mnras] {10.1093/mnras/stz104}, \href
  {https://ui.adsabs.harvard.edu/abs/2019MNRAS.484.3476C} {484, 3476}

\bibitem[\protect\citeauthoryear{{Coccato}, {Morelli}, {Corsini}, {Buson},
  {Pizzella}, {Vergani}  \& {Bertola}}{{Coccato} et~al.}{2011}]{cocatto2011}
{Coccato} L.,  {Morelli} L.,  {Corsini} E.~M.,  {Buson} L.,  {Pizzella} A.,
  {Vergani} D.,   {Bertola} F.,  2011, \mn@doi [\mnras]
  {10.1111/j.1745-3933.2011.01016.x}, \href
  {https://ui.adsabs.harvard.edu/abs/2011MNRAS.412L.113C} {412, L113}

\bibitem[\protect\citeauthoryear{{Coccato}, {Morelli}, {Pizzella}, {Corsini},
  {Buson}  \& {Dalla Bont{\`a}}}{{Coccato} et~al.}{2013}]{cocatto2013}
{Coccato} L.,  {Morelli} L.,  {Pizzella} A.,  {Corsini} E.~M.,  {Buson} L.~M.,
   {Dalla Bont{\`a}} E.,  2013, \mn@doi [\aap] {10.1051/0004-6361/201220460},
  \href {https://ui.adsabs.harvard.edu/abs/2013A&A...549A...3C} {549, A3}

\bibitem[\protect\citeauthoryear{{Cole}, {Debattista}, {Erwin}, {Earp}  \&
  {Ro{\v{s}}kar}}{{Cole} et~al.}{2014}]{cole+14}
{Cole} D.~R.,  {Debattista} V.~P.,  {Erwin} P.,  {Earp} S. W.~F.,
  {Ro{\v{s}}kar} R.,  2014, \mn@doi [\mnras] {10.1093/mnras/stu1985}, \href
  {https://ui.adsabs.harvard.edu/abs/2014MNRAS.445.3352C} {445, 3352}

\bibitem[\protect\citeauthoryear{{Collett}, {Dutta}  \& {Evans}}{{Collett}
  et~al.}{1997}]{collett1997}
{Collett} J.~L.,  {Dutta} S.~N.,   {Evans} N.~W.,  1997, \mn@doi [\mnras]
  {10.1093/mnras/285.1.49}, \href
  {https://ui.adsabs.harvard.edu/abs/1997MNRAS.285...49C} {285, 49}

\bibitem[\protect\citeauthoryear{{Combes}}{{Combes}}{2003}]{combes2003}
{Combes} F.,  2003, in {Combes} F.,  {Barret} D.,  {Contini} T.,   {Pagani} L.,
   eds, SF2A-2003: Semaine de l'Astrophysique Francaise. EDP Sciences, Paris,
  p.~243 (\mn@eprint {arXiv} {astro-ph/0308022})

\bibitem[\protect\citeauthoryear{{Combes} \& {Gerin}}{{Combes} \&
  {Gerin}}{1985}]{combes1985}
{Combes} F.,  {Gerin} M.,  1985, \aap, \href
  {https://ui.adsabs.harvard.edu/abs/1985A&A...150..327C} {150, 327}

\bibitem[\protect\citeauthoryear{{Combes} \& {Sanders}}{{Combes} \&
  {Sanders}}{1981}]{combes1981}
{Combes} F.,  {Sanders} R.~H.,  1981, \aap, \href
  {https://ui.adsabs.harvard.edu/abs/1981A&A....96..164C} {96, 164}

\bibitem[\protect\citeauthoryear{{Combes}, {Debbasch}, {Friedli}  \&
  {Pfenniger}}{{Combes} et~al.}{1990}]{combes1990}
{Combes} F.,  {Debbasch} F.,  {Friedli} D.,   {Pfenniger} D.,  1990, \aap,
  \href {https://ui.adsabs.harvard.edu/abs/1990A&A...233...82C} {233, 82}

\bibitem[\protect\citeauthoryear{{Cowie}, {Hu}  \& {Songaila}}{{Cowie}
  et~al.}{1995}]{cowie1995}
{Cowie} L.~L.,  {Hu} E.~M.,   {Songaila} A.,  1995, \mn@doi [\aj]
  {10.1086/117631}, \href
  {https://ui.adsabs.harvard.edu/abs/1995AJ....110.1576C} {110, 1576}

\bibitem[\protect\citeauthoryear{{Das}, {Hawkins}  \& {Jofr{\'e}}}{{Das}
  et~al.}{2020}]{Das2020}
{Das} P.,  {Hawkins} K.,   {Jofr{\'e}} P.,  2020, \mn@doi [\mnras]
  {10.1093/mnras/stz3537}, \href
  {https://ui.adsabs.harvard.edu/abs/2020MNRAS.493.5195D} {493, 5195}

\bibitem[\protect\citeauthoryear{{Debattista} \& {Sellwood}}{{Debattista} \&
  {Sellwood}}{2000}]{Debattista2000}
{Debattista} V.~P.,  {Sellwood} J.~A.,  2000, \mn@doi [\apj] {10.1086/317148},
  \href {https://ui.adsabs.harvard.edu/abs/2000ApJ...543..704D} {543, 704}

\bibitem[\protect\citeauthoryear{{Debattista}, {Carollo}, {Mayer}  \&
  {Moore}}{{Debattista} et~al.}{2004}]{debattista2004}
{Debattista} V.~P.,  {Carollo} C.~M.,  {Mayer} L.,   {Moore} B.,  2004, \mn@doi
  [\apjl] {10.1086/386332}, \href
  {https://ui.adsabs.harvard.edu/abs/2004ApJ...604L..93D} {604, L93}

\bibitem[\protect\citeauthoryear{{Debattista}, {Mayer}, {Carollo}, {Moore},
  {Wadsley}  \& {Quinn}}{{Debattista} et~al.}{2006}]{Debattista2006}
{Debattista} V.~P.,  {Mayer} L.,  {Carollo} C.~M.,  {Moore} B.,  {Wadsley} J.,
   {Quinn} T.,  2006, \mn@doi [\apj] {10.1086/504147}, \href
  {https://ui.adsabs.harvard.edu/abs/2006ApJ...645..209D} {645, 209}

\bibitem[\protect\citeauthoryear{{Debattista}, {Ness}, {Gonzalez}, {Freeman},
  {Zoccali}  \& {Minniti}}{{Debattista} et~al.}{2017}]{Debattista+17}
{Debattista} V.~P.,  {Ness} M.,  {Gonzalez} O.~A.,  {Freeman} K.,  {Zoccali}
  M.,   {Minniti} D.,  2017, \mn@doi [\mnras] {10.1093/mnras/stx947}, \href
  {https://ui.adsabs.harvard.edu/abs/2017MNRAS.469.1587D} {469, 1587}

\bibitem[\protect\citeauthoryear{{Dekel}, {Sari}  \& {Ceverino}}{{Dekel}
  et~al.}{2009}]{dekel2009}
{Dekel} A.,  {Sari} R.,   {Ceverino} D.,  2009, \mn@doi [\apj]
  {10.1088/0004-637X/703/1/785}, \href
  {https://ui.adsabs.harvard.edu/abs/2009ApJ...703..785D} {703, 785}

\bibitem[\protect\citeauthoryear{{Dessauges-Zavadsky}, {Schaerer}, {Cava},
  {Mayer}  \& {Tamburello}}{{Dessauges-Zavadsky} et~al.}{2017}]{dessauges2017}
{Dessauges-Zavadsky} M.,  {Schaerer} D.,  {Cava} A.,  {Mayer} L.,
  {Tamburello} V.,  2017, \mn@doi [\apjl] {10.3847/2041-8213/aa5d52}, \href
  {https://ui.adsabs.harvard.edu/abs/2017ApJ...836L..22D} {836, L22}

\bibitem[\protect\citeauthoryear{{Elmegreen}}{{Elmegreen}}{2009}]{elmegreen2008}
{Elmegreen} B.~G.,  2009, in {Jogee} S.,  {Marinova} I.,  {Hao} L.,   {Blanc}
  G.~A.,  eds,  Astronomical Society of the Pacific Conference Series Vol. 419,
  Galaxy Evolution: Emerging Insights and Future Challenges. p.~23 (\mn@eprint
  {arXiv} {0903.1937})

\bibitem[\protect\citeauthoryear{{Elmegreen}, {Elmegreen}, {Ravindranath}  \&
  {Coe}}{{Elmegreen} et~al.}{2007}]{elmegreen2007}
{Elmegreen} D.~M.,  {Elmegreen} B.~G.,  {Ravindranath} S.,   {Coe} D.~A.,
  2007, \mn@doi [\apj] {10.1086/511667}, \href
  {https://ui.adsabs.harvard.edu/abs/2007ApJ...658..763E} {658, 763}

\bibitem[\protect\citeauthoryear{{Evans} \& {Collett}}{{Evans} \&
  {Collett}}{1994}]{evans1994}
{Evans} N.~W.,  {Collett} J.~L.,  1994, \mn@doi [\apjl] {10.1086/187164}, \href
  {https://ui.adsabs.harvard.edu/abs/1994ApJ...420L..67E} {420, L67}

\bibitem[\protect\citeauthoryear{{F{\"o}rster Schreiber} et~al.,}{{F{\"o}rster
  Schreiber} et~al.}{2011}]{schreiber2011}
{F{\"o}rster Schreiber} N.~M.,  et~al., 2011, \mn@doi [\apj]
  {10.1088/0004-637X/739/1/45}, \href
  {https://ui.adsabs.harvard.edu/abs/2011ApJ...739...45F} {739, 45}

\bibitem[\protect\citeauthoryear{{Frankel}, {Sanders}, {Rix}, {Ting}  \&
  {Ness}}{{Frankel} et~al.}{2019}]{frankel2019}
{Frankel} N.,  {Sanders} J.,  {Rix} H.-W.,  {Ting} Y.-S.,   {Ness} M.,  2019,
  \mn@doi [\apj] {10.3847/1538-4357/ab4254}, \href
  {https://ui.adsabs.harvard.edu/abs/2019ApJ...884...99F} {884, 99}

\bibitem[\protect\citeauthoryear{{Friedli} \& {Benz}}{{Friedli} \&
  {Benz}}{1993}]{friedli1993}
{Friedli} D.,  {Benz} W.,  1993, \aap, \href
  {https://ui.adsabs.harvard.edu/abs/1993A&A...268...65F} {268, 65}

\bibitem[\protect\citeauthoryear{{Genzel} et~al.,}{{Genzel}
  et~al.}{2011}]{genzel2011}
{Genzel} R.,  et~al., 2011, \mn@doi [\apj] {10.1088/0004-637X/733/2/101}, \href
  {https://ui.adsabs.harvard.edu/abs/2011ApJ...733..101G} {733, 101}

\bibitem[\protect\citeauthoryear{{Ghosh}, {Debattista}  \&
  {Khachaturyants}}{{Ghosh} et~al.}{2020}]{ghosh2020}
{Ghosh} S.,  {Debattista} V.~P.,   {Khachaturyants} T.,  2020, arXiv e-prints,
  \href {https://ui.adsabs.harvard.edu/abs/2020arXiv200902343G} {p.
  arXiv:2009.02343}

\bibitem[\protect\citeauthoryear{{Grand} et~al.,}{{Grand}
  et~al.}{2020}]{grand2020}
{Grand} R. J.~J.,  et~al., 2020, \mn@doi [\mnras] {10.1093/mnras/staa2057},
  \href {https://ui.adsabs.harvard.edu/abs/2020MNRAS.497.1603G} {497, 1603}

\bibitem[\protect\citeauthoryear{{Guo}, {Giavalisco}, {Ferguson}, {Cassata}  \&
  {Koekemoer}}{{Guo} et~al.}{2012}]{guo2012}
{Guo} Y.,  {Giavalisco} M.,  {Ferguson} H.~C.,  {Cassata} P.,   {Koekemoer}
  A.~M.,  2012, \mn@doi [\apj] {10.1088/0004-637X/757/2/120}, \href
  {https://ui.adsabs.harvard.edu/abs/2012ApJ...757..120G} {757, 120}

\bibitem[\protect\citeauthoryear{{Guo} et~al.,}{{Guo} et~al.}{2015}]{guo2015}
{Guo} Y.,  et~al., 2015, \mn@doi [\apj] {10.1088/0004-637X/800/1/39}, \href
  {https://ui.adsabs.harvard.edu/abs/2015ApJ...800...39G} {800, 39}

\bibitem[\protect\citeauthoryear{{Hasan} \& {Norman}}{{Hasan} \&
  {Norman}}{1990}]{hasan1990}
{Hasan} H.,  {Norman} C.,  1990, \mn@doi [\apj] {10.1086/169168}, \href
  {https://ui.adsabs.harvard.edu/abs/1990ApJ...361...69H} {361, 69}

\bibitem[\protect\citeauthoryear{{Hawkins}, {Jofr{\'e}}, {Masseron}  \&
  {Gilmore}}{{Hawkins} et~al.}{2015}]{hawkins2015}
{Hawkins} K.,  {Jofr{\'e}} P.,  {Masseron} T.,   {Gilmore} G.,  2015, \mn@doi
  [\mnras] {10.1093/mnras/stv1586}, \href
  {https://ui.adsabs.harvard.edu/abs/2015MNRAS.453..758H} {453, 758}

\bibitem[\protect\citeauthoryear{{Hayes} et~al.,}{{Hayes}
  et~al.}{2018}]{hayes2018}
{Hayes} C.~R.,  et~al., 2018, \mn@doi [\apj] {10.3847/1538-4357/aa9cec}, \href
  {https://ui.adsabs.harvard.edu/\#abs/2018ApJ...852...49H} {852, 49}

\bibitem[\protect\citeauthoryear{{Helmi}, {Babusiaux}, {Koppelman}, {Massari},
  {Veljanoski}  \& {Brown}}{{Helmi} et~al.}{2018}]{helmi2018}
{Helmi} A.,  {Babusiaux} C.,  {Koppelman} H.~H.,  {Massari} D.,  {Veljanoski}
  J.,   {Brown} A. G.~A.,  2018, \mn@doi [\nat] {10.1038/s41586-018-0625-x},
  \href {https://ui.adsabs.harvard.edu/abs/2018Natur.563...85H} {563, 85}

\bibitem[\protect\citeauthoryear{{Hernquist} \& {Weinberg}}{{Hernquist} \&
  {Weinberg}}{1992}]{Hernquist1992}
{Hernquist} L.,  {Weinberg} M.~D.,  1992, \mn@doi [\apj] {10.1086/171975},
  \href {https://ui.adsabs.harvard.edu/abs/1992ApJ...400...80H} {400, 80}

\bibitem[\protect\citeauthoryear{{Herpich}, {Stinson}, {Rix}, {Martig}  \&
  {Dutton}}{{Herpich} et~al.}{2017}]{Herpich2017}
{Herpich} J.,  {Stinson} G.~S.,  {Rix} H.~W.,  {Martig} M.,   {Dutton} A.~A.,
  2017, \mn@doi [\mnras] {10.1093/mnras/stx1511}, \href
  {https://ui.adsabs.harvard.edu/abs/2017MNRAS.470.4941H} {470, 4941}

\bibitem[\protect\citeauthoryear{{Ho}}{{Ho}}{2008}]{ho2008}
{Ho} L.~C.,  2008, \mn@doi [\araa] {10.1146/annurev.astro.45.051806.110546},
  \href {https://ui.adsabs.harvard.edu/abs/2008ARA&A..46..475H} {46, 475}

\bibitem[\protect\citeauthoryear{{Hohl}}{{Hohl}}{1971}]{hohl1971}
{Hohl} F.,  1971, \mn@doi [\apj] {10.1086/151091}, \href
  {https://ui.adsabs.harvard.edu/abs/1971ApJ...168..343H} {168, 343}

\bibitem[\protect\citeauthoryear{{Hohl}}{{Hohl}}{1978}]{hohl1978}
{Hohl} F.,  1978, \mn@doi [\aj] {10.1086/112259}, \href
  {https://ui.adsabs.harvard.edu/abs/1978AJ.....83..768H} {83, 768}

\bibitem[\protect\citeauthoryear{{Johnston}, {Merrifield},
  {Arag{\'o}n-Salamanca}  \& {Cappellari}}{{Johnston}
  et~al.}{2013}]{johnston2013}
{Johnston} E.~J.,  {Merrifield} M.~R.,  {Arag{\'o}n-Salamanca} A.,
  {Cappellari} M.,  2013, \mn@doi [\mnras] {10.1093/mnras/sts121}, \href
  {https://ui.adsabs.harvard.edu/abs/2013MNRAS.428.1296J} {428, 1296}

\bibitem[\protect\citeauthoryear{{Kalnajs}}{{Kalnajs}}{1973}]{Kalnajs1973}
{Kalnajs} A.~J.,  1973, \mn@doi [Proceedings of the Astronomical Society of
  Australia] {10.1017/S1323358000013461}, \href
  {https://ui.adsabs.harvard.edu/abs/1973PASAu...2..174K} {2, 174}

\bibitem[\protect\citeauthoryear{{Katkov}, {Sil'chenko}  \&
  {Afanasiev}}{{Katkov} et~al.}{2013}]{katkov2013}
{Katkov} I.~Y.,  {Sil'chenko} O.~K.,   {Afanasiev} V.~L.,  2013, \mn@doi [\apj]
  {10.1088/0004-637X/769/2/105}, \href
  {https://ui.adsabs.harvard.edu/abs/2013ApJ...769..105K} {769, 105}

\bibitem[\protect\citeauthoryear{{Kazantzidis}, {Zentner}, {Kravtsov},
  {Bullock}  \& {Debattista}}{{Kazantzidis} et~al.}{2009}]{kazandzitis2009}
{Kazantzidis} S.,  {Zentner} A.~R.,  {Kravtsov} A.~V.,  {Bullock} J.~S.,
  {Debattista} V.~P.,  2009, \mn@doi [\apj] {10.1088/0004-637X/700/2/1896},
  \href {https://ui.adsabs.harvard.edu/abs/2009ApJ...700.1896K} {700, 1896}

\bibitem[\protect\citeauthoryear{{Kordopatis} et~al.,}{{Kordopatis}
  et~al.}{2013}]{kordopatis2013}
{Kordopatis} G.,  et~al., 2013, \mn@doi [\mnras] {10.1093/mnras/stt1804}, \href
  {https://ui.adsabs.harvard.edu/abs/2013MNRAS.436.3231K} {436, 3231}

\bibitem[\protect\citeauthoryear{{Kordopatis}, {Recio-Blanco}, {Schultheis}  \&
  {Hill}}{{Kordopatis} et~al.}{2020}]{kordopatis2020}
{Kordopatis} G.,  {Recio-Blanco} A.,  {Schultheis} M.,   {Hill} V.,  2020,
  \mn@doi [\aap] {10.1051/0004-6361/202038686}, \href
  {https://ui.adsabs.harvard.edu/abs/2020A&A...643A..69K} {643, A69}

\bibitem[\protect\citeauthoryear{{Kormendy}}{{Kormendy}}{2013}]{Kormendy2013}
{Kormendy} J.,  2013, {Secular Evolution in Disk Galaxies}.
Cambridge Univ. Press, Cambridge, p.~1

\bibitem[\protect\citeauthoryear{{Kormendy} \& {Kennicutt}}{{Kormendy} \&
  {Kennicutt}}{2004}]{Kormendy2004}
{Kormendy} J.,  {Kennicutt} Robert~C. J.,  2004, \mn@doi [\araa]
  {10.1146/annurev.astro.42.053102.134024}, \href
  {https://ui.adsabs.harvard.edu/abs/2004ARA&A..42..603K} {42, 603}

\bibitem[\protect\citeauthoryear{{Kuijken}, {Fisher}  \&
  {Merrifield}}{{Kuijken} et~al.}{1996}]{kuijken1996}
{Kuijken} K.,  {Fisher} D.,   {Merrifield} M.~R.,  1996, \mn@doi [\mnras]
  {10.1093/mnras/283.2.543}, \href
  {https://ui.adsabs.harvard.edu/abs/1996MNRAS.283..543K} {283, 543}

\bibitem[\protect\citeauthoryear{{Martinez-Valpuesta}, {Shlosman}  \&
  {Heller}}{{Martinez-Valpuesta} et~al.}{2006}]{valpuesta2006}
{Martinez-Valpuesta} I.,  {Shlosman} I.,   {Heller} C.,  2006, \mn@doi [\apj]
  {10.1086/498338}, \href
  {https://ui.adsabs.harvard.edu/abs/2006ApJ...637..214M} {637, 214}

\bibitem[\protect\citeauthoryear{{Merrifield} \& {Kuijken}}{{Merrifield} \&
  {Kuijken}}{1994}]{merrifield1994}
{Merrifield} M.~R.,  {Kuijken} K.,  1994, \mn@doi [\apj] {10.1086/174596},
  \href {https://ui.adsabs.harvard.edu/abs/1994ApJ...432..575M} {432, 575}

\bibitem[\protect\citeauthoryear{{Merritt} \& {Sellwood}}{{Merritt} \&
  {Sellwood}}{1994}]{merritt1994}
{Merritt} D.,  {Sellwood} J.~A.,  1994, \mn@doi [\apj] {10.1086/174005}, \href
  {https://ui.adsabs.harvard.edu/abs/1994ApJ...425..551M} {425, 551}

\bibitem[\protect\citeauthoryear{{Minchev}, {Famaey}, {Quillen}, {Di Matteo},
  {Combes}, {Vlaji{\'c}}, {Erwin}  \& {Bland -Hawthorn}}{{Minchev}
  et~al.}{2012}]{Minchev2012}
{Minchev} I.,  {Famaey} B.,  {Quillen} A.~C.,  {Di Matteo} P.,  {Combes} F.,
  {Vlaji{\'c}} M.,  {Erwin} P.,   {Bland -Hawthorn} J.,  2012, \mn@doi [\aap]
  {10.1051/0004-6361/201219198}, \href
  {https://ui.adsabs.harvard.edu/abs/2012A&A...548A.126M} {548, A126}

\bibitem[\protect\citeauthoryear{{Moetazedian} \& {Just}}{{Moetazedian} \&
  {Just}}{2016}]{moetezadian2016}
{Moetazedian} R.,  {Just} A.,  2016, \mn@doi [\mnras] {10.1093/mnras/stw764},
  \href {https://ui.adsabs.harvard.edu/abs/2016MNRAS.459.2905M} {459, 2905}

\bibitem[\protect\citeauthoryear{{Mu{\~n}oz-Mateos}, {Gil de Paz}, {Boissier},
  {Zamorano}, {Jarrett}, {Gallego}  \& {Madore}}{{Mu{\~n}oz-Mateos}
  et~al.}{2007}]{munoz2007}
{Mu{\~n}oz-Mateos} J.~C.,  {Gil de Paz} A.,  {Boissier} S.,  {Zamorano} J.,
  {Jarrett} T.,  {Gallego} J.,   {Madore} B.~F.,  2007, \mn@doi [\apj]
  {10.1086/511812}, \href
  {https://ui.adsabs.harvard.edu/abs/2007ApJ...658.1006M} {658, 1006}

\bibitem[\protect\citeauthoryear{{Navarro}, {Frenk}  \& {White}}{{Navarro}
  et~al.}{1997}]{NFW}
{Navarro} J.~F.,  {Frenk} C.~S.,   {White} S. D.~M.,  1997, \mn@doi [\apj]
  {10.1086/304888}, \href
  {https://ui.adsabs.harvard.edu/abs/1997ApJ...490..493N} {490, 493}

\bibitem[\protect\citeauthoryear{{Noguchi}}{{Noguchi}}{1999}]{noguchi1999}
{Noguchi} M.,  1999, \mn@doi [\apj] {10.1086/306932}, \href
  {https://ui.adsabs.harvard.edu/abs/1999ApJ...514...77N} {514, 77}

\bibitem[\protect\citeauthoryear{{Overzier}, {Heckman}, {Schiminovich},
  {Basu-Zych}, {Gon{\c{c}}alves}, {Martin}  \& {Rich}}{{Overzier}
  et~al.}{2010}]{overzier2010}
{Overzier} R.~A.,  {Heckman} T.~M.,  {Schiminovich} D.,  {Basu-Zych} A.,
  {Gon{\c{c}}alves} T.,  {Martin} D.~C.,   {Rich} R.~M.,  2010, \mn@doi [\apj]
  {10.1088/0004-637X/710/2/979}, \href
  {https://ui.adsabs.harvard.edu/abs/2010ApJ...710..979O} {710, 979}

\bibitem[\protect\citeauthoryear{{Peebles}}{{Peebles}}{1969}]{peebles1969}
{Peebles} P.~J.~E.,  1969, \mn@doi [\apj] {10.1086/149876}, \href
  {https://ui.adsabs.harvard.edu/abs/1969ApJ...155..393P} {155, 393}

\bibitem[\protect\citeauthoryear{{Pfenniger}}{{Pfenniger}}{1984}]{pfenniger1984}
{Pfenniger} D.,  1984, \aap, \href
  {https://ui.adsabs.harvard.edu/abs/1984A&A...134..373P} {134, 373}

\bibitem[\protect\citeauthoryear{{Pfenniger} \& {Norman}}{{Pfenniger} \&
  {Norman}}{1990}]{pfenniger1990}
{Pfenniger} D.,  {Norman} C.,  1990, \mn@doi [\apj] {10.1086/169352}, \href
  {https://ui.adsabs.harvard.edu/abs/1990ApJ...363..391P} {363, 391}

\bibitem[\protect\citeauthoryear{{Pizzella}, {Morelli}, {Corsini}, {Dalla
  Bont{\`a}}, {Coccato}  \& {Sanjana}}{{Pizzella} et~al.}{2014}]{pizzella2014}
{Pizzella} A.,  {Morelli} L.,  {Corsini} E.~M.,  {Dalla Bont{\`a}} E.,
  {Coccato} L.,   {Sanjana} G.,  2014, \mn@doi [\aap]
  {10.1051/0004-6361/201424746}, \href
  {https://ui.adsabs.harvard.edu/abs/2014A&A...570A..79P} {570, A79}

\bibitem[\protect\citeauthoryear{{Pizzella}, {Morelli}, {Coccato}, {Corsini},
  {Dalla Bont{\`a}}, {Fabricius}  \& {Saglia}}{{Pizzella}
  et~al.}{2018}]{pizzella2018}
{Pizzella} A.,  {Morelli} L.,  {Coccato} L.,  {Corsini} E.~M.,  {Dalla
  Bont{\`a}} E.,  {Fabricius} M.,   {Saglia} R.~P.,  2018, \mn@doi [\aap]
  {10.1051/0004-6361/201731712}, \href
  {https://ui.adsabs.harvard.edu/abs/2018A&A...616A..22P} {616, A22}

\bibitem[\protect\citeauthoryear{{Pontzen}, {Ro{\v s}kar}, {Stinson}, {Woods},
  {Reed}, {Coles}  \& {Quinn}}{{Pontzen} et~al.}{2013}]{pynbody}
{Pontzen} A.,  {Ro{\v s}kar} R.,  {Stinson} G.~S.,  {Woods} R.,  {Reed} D.~M.,
  {Coles} J.,   {Quinn} T.~R.,  2013, {pynbody: Astrophysics Simulation
  Analysis for Python}

\bibitem[\protect\citeauthoryear{{Portaluri} et~al.,}{{Portaluri}
  et~al.}{2017}]{portaluri2017}
{Portaluri} E.,  et~al., 2017, \mn@doi [\mnras] {10.1093/mnras/stx172}, \href
  {https://ui.adsabs.harvard.edu/abs/2017MNRAS.467.1008P} {467, 1008}

\bibitem[\protect\citeauthoryear{{Posti}, {Helmi}, {Veljanoski}  \&
  {Breddels}}{{Posti} et~al.}{2018}]{posti2018}
{Posti} L.,  {Helmi} A.,  {Veljanoski} J.,   {Breddels} M.~A.,  2018, \mn@doi
  [\aap] {10.1051/0004-6361/201732277}, \href
  {https://ui.adsabs.harvard.edu/abs/2018A&A...615A..70P} {615, A70}

\bibitem[\protect\citeauthoryear{{Puerari} \& {Pfenniger}}{{Puerari} \&
  {Pfenniger}}{2001}]{puerari2001}
{Puerari} I.,  {Pfenniger} D.,  2001, \mn@doi [\apss]
  {10.1023/A:1017581325673}, \href
  {https://ui.adsabs.harvard.edu/abs/2001Ap&SS.276..909P} {276, 909}

\bibitem[\protect\citeauthoryear{{Queiroz} et~al.,}{{Queiroz}
  et~al.}{2020}]{queiroz2020}
{Queiroz} A.~B.~A.,  et~al., 2020, arXiv e-prints, \href
  {https://ui.adsabs.harvard.edu/abs/2020arXiv200712915Q} {p. arXiv:2007.12915}

\bibitem[\protect\citeauthoryear{{Raha}, {Sellwood}, {James}  \& {Kahn}}{{Raha}
  et~al.}{1991}]{raha1991}
{Raha} N.,  {Sellwood} J.~A.,  {James} R.~A.,   {Kahn} F.~D.,  1991, \mn@doi
  [\nat] {10.1038/352411a0}, \href
  {https://ui.adsabs.harvard.edu/abs/1991Natur.352..411R} {352, 411}

\bibitem[\protect\citeauthoryear{{Ravindranath} et~al.,}{{Ravindranath}
  et~al.}{2006}]{ravindranath2006}
{Ravindranath} S.,  et~al., 2006, \mn@doi [\apj] {10.1086/507016}, \href
  {https://ui.adsabs.harvard.edu/abs/2006ApJ...652..963R} {652, 963}

\bibitem[\protect\citeauthoryear{{Rix}, {Franx}, {Fisher}  \&
  {Illingworth}}{{Rix} et~al.}{1992}]{rix1992}
{Rix} H.-W.,  {Franx} M.,  {Fisher} D.,   {Illingworth} G.,  1992, \mn@doi
  [\apjl] {10.1086/186635}, \href
  {https://ui.adsabs.harvard.edu/abs/1992ApJ...400L...5R} {400, L5}

\bibitem[\protect\citeauthoryear{{Ro{\v{s}}kar}, {Debattista}, {Quinn},
  {Stinson}  \& {Wadsley}}{{Ro{\v{s}}kar} et~al.}{2008}]{roskar2008}
{Ro{\v{s}}kar} R.,  {Debattista} V.~P.,  {Quinn} T.~R.,  {Stinson} G.~S.,
  {Wadsley} J.,  2008, \mn@doi [\apjl] {10.1086/592231}, \href
  {https://ui.adsabs.harvard.edu/abs/2008ApJ...684L..79R} {684, L79}

\bibitem[\protect\citeauthoryear{{Rubin}, {Graham}  \& {Kenney}}{{Rubin}
  et~al.}{1992}]{rubin1992}
{Rubin} V.~C.,  {Graham} J.~A.,   {Kenney} J. D.~P.,  1992, \mn@doi [\apjl]
  {10.1086/186460}, \href
  {https://ui.adsabs.harvard.edu/abs/1992ApJ...394L...9R} {394, L9}

\bibitem[\protect\citeauthoryear{{Sakamoto}, {Okumura}, {Ishizuki}  \&
  {Scoville}}{{Sakamoto} et~al.}{1999}]{sakamoto1999}
{Sakamoto} K.,  {Okumura} S.~K.,  {Ishizuki} S.,   {Scoville} N.~Z.,  1999,
  \mn@doi [\apjs] {10.1086/313265}, \href
  {https://ui.adsabs.harvard.edu/abs/1999ApJS..124..403S} {124, 403}

\bibitem[\protect\citeauthoryear{{S{\'a}nchez} et~al.,}{{S{\'a}nchez}
  et~al.}{2012}]{sanchez2012}
{S{\'a}nchez} S.~F.,  et~al., 2012, \mn@doi [\aap]
  {10.1051/0004-6361/201117353}, \href
  {https://ui.adsabs.harvard.edu/abs/2012A&A...538A...8S} {538, A8}

\bibitem[\protect\citeauthoryear{{Sellwood}}{{Sellwood}}{1980}]{sellwood1980}
{Sellwood} J.~A.,  1980, \aap, \href
  {https://ui.adsabs.harvard.edu/abs/1980A&A....89..296S} {89, 296}

\bibitem[\protect\citeauthoryear{{Sellwood} \& {Athanassoula}}{{Sellwood} \&
  {Athanassoula}}{1986}]{Sellwood1986}
{Sellwood} J.~A.,  {Athanassoula} E.,  1986, \mn@doi [\mnras]
  {10.1093/mnras/221.2.195}, \href
  {https://ui.adsabs.harvard.edu/abs/1986MNRAS.221..195S} {221, 195}

\bibitem[\protect\citeauthoryear{{Sharma}, {Hayden}  \&
  {Bland-Hawthorn}}{{Sharma} et~al.}{2020}]{sharma2020}
{Sharma} S.,  {Hayden} M.~R.,   {Bland-Hawthorn} J.,  2020, arXiv e-prints,
  \href {https://ui.adsabs.harvard.edu/abs/2020arXiv200503646S} {p.
  arXiv:2005.03646}

\bibitem[\protect\citeauthoryear{{Shen} \& {Sellwood}}{{Shen} \&
  {Sellwood}}{2004}]{shen2004}
{Shen} J.,  {Sellwood} J.~A.,  2004, \mn@doi [\apj] {10.1086/382124}, \href
  {https://ui.adsabs.harvard.edu/abs/2004ApJ...604..614S} {604, 614}

\bibitem[\protect\citeauthoryear{{Shen}, {Wadsley}  \& {Stinson}}{{Shen}
  et~al.}{2010}]{ShenWadsleyStinson10}
{Shen} S.,  {Wadsley} J.,   {Stinson} G.,  2010, \mn@doi [\mnras]
  {10.1111/j.1365-2966.2010.17047.x}, \href
  {https://ui.adsabs.harvard.edu/abs/2010MNRAS.407.1581S} {407, 1581}

\bibitem[\protect\citeauthoryear{{Sheth}, {Vogel}, {Regan}, {Thornley}  \&
  {Teuben}}{{Sheth} et~al.}{2005}]{sheth2005}
{Sheth} K.,  {Vogel} S.~N.,  {Regan} M.~W.,  {Thornley} M.~D.,   {Teuben}
  P.~J.,  2005, \mn@doi [\apj] {10.1086/432409}, \href
  {https://ui.adsabs.harvard.edu/abs/2005ApJ...632..217S} {632, 217}

\bibitem[\protect\citeauthoryear{{Soto} et~al.,}{{Soto}
  et~al.}{2017}]{soto2017}
{Soto} E.,  et~al., 2017, \mn@doi [\apj] {10.3847/1538-4357/aa5da3}, \href
  {https://ui.adsabs.harvard.edu/abs/2017ApJ...837....6S} {837, 6}

\bibitem[\protect\citeauthoryear{{Sparke} \& {Sellwood}}{{Sparke} \&
  {Sellwood}}{1987}]{sparke1987}
{Sparke} L.~S.,  {Sellwood} J.~A.,  1987, \mn@doi [\mnras]
  {10.1093/mnras/225.3.653}, \href
  {https://ui.adsabs.harvard.edu/abs/1987MNRAS.225..653S} {225, 653}

\bibitem[\protect\citeauthoryear{{Stinson}, {Seth}, {Katz}, {Wadsley},
  {Governato}  \& {Quinn}}{{Stinson} et~al.}{2006}]{stinson2006}
{Stinson} G.,  {Seth} A.,  {Katz} N.,  {Wadsley} J.,  {Governato} F.,   {Quinn}
  T.,  2006, \mn@doi [\mnras] {10.1111/j.1365-2966.2006.11097.x}, \href
  {https://ui.adsabs.harvard.edu/abs/2006MNRAS.373.1074S} {373, 1074}

\bibitem[\protect\citeauthoryear{{Swinbank} et~al.,}{{Swinbank}
  et~al.}{2010}]{swinbank2010}
{Swinbank} A.~M.,  et~al., 2010, \mn@doi [\mnras]
  {10.1111/j.1365-2966.2010.16485.x}, \href
  {https://ui.adsabs.harvard.edu/abs/2010MNRAS.405..234S} {405, 234}

\bibitem[\protect\citeauthoryear{{Thakar} \& {Ryden}}{{Thakar} \&
  {Ryden}}{1996}]{thakar1996}
{Thakar} A.~R.,  {Ryden} B.~S.,  1996, \mn@doi [\apj] {10.1086/177037}, \href
  {https://ui.adsabs.harvard.edu/abs/1996ApJ...461...55T} {461, 55}

\bibitem[\protect\citeauthoryear{{Thakar} \& {Ryden}}{{Thakar} \&
  {Ryden}}{1998}]{thakar1998}
{Thakar} A.~R.,  {Ryden} B.~S.,  1998, \mn@doi [\apj] {10.1086/306223}, \href
  {https://ui.adsabs.harvard.edu/abs/1998ApJ...506...93T} {506, 93}

\bibitem[\protect\citeauthoryear{{Toth} \& {Ostriker}}{{Toth} \&
  {Ostriker}}{1992}]{toth1992}
{Toth} G.,  {Ostriker} J.~P.,  1992, \mn@doi [\apj] {10.1086/171185}, \href
  {https://ui.adsabs.harvard.edu/abs/1992ApJ...389....5T} {389, 5}

\bibitem[\protect\citeauthoryear{{Tremaine} \& {Weinberg}}{{Tremaine} \&
  {Weinberg}}{1984}]{Tremaine1984}
{Tremaine} S.,  {Weinberg} M.~D.,  1984, \mn@doi [\mnras]
  {10.1093/mnras/209.4.729}, \href
  {https://ui.adsabs.harvard.edu/abs/1984MNRAS.209..729T} {209, 729}

\bibitem[\protect\citeauthoryear{{Valenzuela} \& {Klypin}}{{Valenzuela} \&
  {Klypin}}{2003}]{Valenzuela2003}
{Valenzuela} O.,  {Klypin} A.,  2003, \mn@doi [\mnras]
  {10.1046/j.1365-8711.2003.06947.x}, \href
  {https://ui.adsabs.harvard.edu/abs/2003MNRAS.345..406V} {345, 406}

\bibitem[\protect\citeauthoryear{{Velazquez} \& {White}}{{Velazquez} \&
  {White}}{1999}]{velazquez1999}
{Velazquez} H.,  {White} S. D.~M.,  1999, \mn@doi [\mnras]
  {10.1046/j.1365-8711.1999.02354.x}, \href
  {https://ui.adsabs.harvard.edu/abs/1999MNRAS.304..254V} {304, 254}

\bibitem[\protect\citeauthoryear{Wadsley, Stadel  \& Quinn}{Wadsley
  et~al.}{2004}]{Wadsleyetal2004}
Wadsley J.,  Stadel J.,   Quinn T.,  2004, \mn@doi [New Astronomy]
  {https://doi.org/10.1016/j.newast.2003.08.004}, 9, 137

\bibitem[\protect\citeauthoryear{{Wadsley}, {Keller}  \& {Quinn}}{{Wadsley}
  et~al.}{2017}]{Wadsley2017}
{Wadsley} J.~W.,  {Keller} B.~W.,   {Quinn} T.~R.,  2017, \mn@doi [\mnras]
  {10.1093/mnras/stx1643}, \href
  {https://ui.adsabs.harvard.edu/abs/2017MNRAS.471.2357W} {471, 2357}

\bibitem[\protect\citeauthoryear{{Weinberg}}{{Weinberg}}{1985}]{Weinberg1985}
{Weinberg} M.~D.,  1985, \mn@doi [\mnras] {10.1093/mnras/213.3.451}, \href
  {https://ui.adsabs.harvard.edu/abs/1985MNRAS.213..451W} {213, 451}

\bibitem[\protect\citeauthoryear{{White} \& {Frenk}}{{White} \&
  {Frenk}}{1991}]{white1991}
{White} S. D.~M.,  {Frenk} C.~S.,  1991, \mn@doi [\apj] {10.1086/170483}, \href
  {https://ui.adsabs.harvard.edu/abs/1991ApJ...379...52W} {379, 52}

\bibitem[\protect\citeauthoryear{{Wozniak} \& {Pfenniger}}{{Wozniak} \&
  {Pfenniger}}{1997}]{wozniak1997}
{Wozniak} H.,  {Pfenniger} D.,  1997, \aap, \href
  {https://ui.adsabs.harvard.edu/abs/1997A&A...317...14W} {317, 14}

\bibitem[\protect\citeauthoryear{{Zeilinger}, {Vega Beltr{\'a}n}, {Rozas},
  {Beckman}, {Pizzella}, {Corsini}  \& {Bertola}}{{Zeilinger}
  et~al.}{2001}]{zeilinger2001}
{Zeilinger} W.~W.,  {Vega Beltr{\'a}n} J.~C.,  {Rozas} M.,  {Beckman} J.~E.,
  {Pizzella} A.,  {Corsini} E.~M.,   {Bertola} F.,  2001, \mn@doi [\apss]
  {10.1023/A:1017548101623}, \href
  {https://ui.adsabs.harvard.edu/abs/2001Ap&SS.276..643Z} {276, 643}

\bibitem[\protect\citeauthoryear{{Zhu} et~al.,}{{Zhu} et~al.}{2018}]{zhu2018}
{Zhu} L.,  et~al., 2018, \mn@doi [Nature Astronomy]
  {10.1038/s41550-017-0348-1}, \href
  {https://ui.adsabs.harvard.edu/abs/2018NatAs...2..233Z} {2, 233}

\bibitem[\protect\citeauthoryear{{van den Bergh}, {Abraham}, {Ellis}, {Tanvir},
  {Santiago}  \& {Glazebrook}}{{van den Bergh} et~al.}{1996}]{van1996}
{van den Bergh} S.,  {Abraham} R.~G.,  {Ellis} R.~S.,  {Tanvir} N.~R.,
  {Santiago} B.~X.,   {Glazebrook} K.~G.,  1996, \mn@doi [\aj]
  {10.1086/118020}, \href
  {https://ui.adsabs.harvard.edu/abs/1996AJ....112..359V} {112, 359}

\makeatother
\end{thebibliography}








\bsp	
\label{lastpage}
\end{document}